\documentclass[journal]{IEEEtran}

\usepackage[T1]{fontenc}
\usepackage[utf8]{inputenc}
\usepackage{graphicx}
\graphicspath{{figures/assets/}}
\usepackage{amsmath,amssymb,amsfonts}
\usepackage{amsthm}
\usepackage{mathrsfs}
\usepackage{textcomp}
\usepackage{xcolor}
\usepackage{booktabs}
\usepackage{multirow}
\usepackage{makecell}
\usepackage{cite}
\usepackage[hidelinks]{hyperref}
\usepackage{url}
\usepackage{caption}
\usepackage{subcaption}
\usepackage{placeins}

\begin{document}
\title{Reliable Grid Forecasting: State Space Models for Safety-Critical Energy Systems}

\author{Sunki~Hong and Jisoo~Lee,~\IEEEmembership{Member,~IEEE}
\thanks{S. Hong and J. Lee are with Gramm AI (e-mail: sunki@gramm.ai; jisoo@gramm.ai).}
\thanks{Corresponding author: Jisoo Lee.}}

\maketitle
\begin{abstract}
Accurate grid load forecasting is safety-critical: under-predictions risk supply shortfalls, while symmetric error metrics can mask this operational asymmetry. We introduce an operator-legible evaluation framework---Under-Prediction Rate (UPR), tail Reserve$_{99.5}^{\%}$ requirements, and explicit inflation diagnostics (Bias$_{24h}$/OPR)---to quantify one-sided reliability risk beyond MAPE.

Using this framework, we evaluate five neural architectures---two state space models (S-Mamba, PowerMamba), two Transformers (iTransformer, PatchTST), an LSTM, and a probabilistic SSM variant (Mamba-ProbTSF)---on a weather-aligned California Independent System Operator (CAISO) dataset spanning Nov 2023--Nov 2025 (84{,}498 hourly records across 5 regional transmission areas) under a rolling-origin walk-forward backtest. We develop and evaluate thermal-lag-aligned weather fusion strategies matched to each architecture's inductive bias.

Our results demonstrate that standard accuracy metrics are insufficient proxies for operational safety: models with comparable MAPE can imply materially different tail reserve requirements (Reserve$_{99.5}^{\%}$). We show that explicit weather integration narrows error distributions, with the magnitude of improvement being architecturally determined---iTransformer's cross-variate attention benefits significantly more than PatchTST's channel-independent design. Crucially, we identify a widespread susceptibility to ``fake safety'' in risk-averse forecasting: while probabilistic calibration reduces upper-tail errors, it achieves this by systematically inflating schedules (e.g., increasing bias by over 1{,}700~MW in severe cases) if left unconstrained. To solve this, we introduce Bias/OPR-constrained objectives that enable auditable trade-offs between minimizing tail risk and preventing trivial over-forecasting.
\end{abstract}


\section{Introduction}

Short-term load forecasting (STLF) underpins electricity grid operations, governing unit commitment, economic dispatch, and reserve scheduling \cite{hong2016probabilistic}. The consequences are significant: under-prediction risks cascading blackouts, while over-prediction incurs unnecessary costs and emissions.

California's grid exemplifies the challenges facing high-renewable systems. Utility-scale solar and wind contribute roughly one quarter of CAISO system energy \cite{caiso2022marketissues}, with behind-the-meter (BTM) solar capacity exceeding 17~GW in California \cite{eia_epm_table_6_02_b}. This invisible generation creates the ``duck curve''---deep midday net load trough followed by steep evening ramps---that evolves annually as BTM capacity expands. Compounding this non-stationarity, climate-driven extreme weather events increasingly trigger non-linear demand spikes that defy historical patterns \cite{caiso2021root}.

Existing forecasting methods face fundamental limitations. Statistical approaches (ARIMA) cannot capture non-linear weather dependencies \cite{eren2024systematic}. Recurrent networks (LSTMs) struggle with long-range dependencies due to vanishing gradients. Transformer architectures address this but introduce quadratic $O(n^2)$ complexity, limiting practical context lengths for capturing multi-week seasonal patterns \cite{vaswani2017attention}.

State space models (SSMs) offer a compelling path to reliability. By achieving linear $O(n)$ scaling, Mamba \cite{gu2023mamba} allows for extended historical contexts and lower inference latency, freeing computational budget for robust uncertainty quantification and ensemble methods. However, Mamba's ability to maintain safety margins in regional grid forecasting---characterized by asymmetric error costs and heteroscedasticity---remains unexplored.

We present a systematic evaluation of state space model and Transformer architectures for reliable California grid load forecasting. Benchmarking Mamba variants against PatchTST, iTransformer, LSTM, and the foundation model Chronos, we shift the evaluation focus from pure accuracy to \textit{operational safety}. Our walk-forward results highlight that accuracy alone is not a sufficient proxy for operational risk: Reserve$_{99.5}^{\%}$ can vary substantially across models even when MAPE is comparable. Furthermore, we demonstrate that explicit weather integration is critical for safety by narrowing error distributions and reducing extreme error events. Crucially, our loss-function ablations reveal a latent danger in risk-averse forecasting: unconstrained probabilistic training achieves apparent reductions in tail risk by systematically inflating schedules (e.g., driving positive bias from +109~MW to +1{,}862~MW). We demonstrate how to prevent this ``fake safety'' using explicit operational constraints.

\subsection*{Contributions}
To move beyond symmetric accuracy metrics toward operator-relevant reliability for California grid forecasting, we make the following contributions:
\begin{enumerate}
    \item \textbf{Grid-Specific Evaluation Framework.} We formalize operational risk metrics (Under-Prediction Rate, tail Reserve$_{99.5}^{\%}$ requirements, Bias/OPR diagnostics) that capture asymmetric operational costs and prevent ``fake safety'' via systematic forecast inflation (Section~\ref{sec:metrics}).

    \item \textbf{Weather Integration for SSMs.} We develop (Section~\ref{sec:weather_arch}) and systematically evaluate (Section~\ref{sec:weather}) thermal-lag-aligned weather fusion strategies for Mamba architectures, improving robustness by narrowing error distributions and reducing extreme errors.

    \item \textbf{Exposing ``Fake Safety'' via Constrained Objectives.} We demonstrate that naive probabilistic calibration (multi-quantile pinball) can trivially reduce large-error events by introducing massive systematic inflation. We introduce and evaluate Bias/OPR constraints to enforce auditable trade-offs between true tail-risk reduction and schedule inflation (Section~\ref{sec:results}).
\end{enumerate}

\section{Background}

\subsection{State Space Models}
State space models (SSMs) provide a principled framework for modeling sequential data through continuous-time dynamical systems. The general linear SSM is defined by the following ordinary differential equations:
\begin{equation}
h'(t) = Ah(t) + Bx(t), \quad y(t) = Ch(t) + Dx(t)
\end{equation}
where $x(t) \in \mathbb{R}$ is the input signal, $h(t) \in \mathbb{R}^N$ is the latent state with dimension $N$, and $y(t) \in \mathbb{R}$ is the output. The matrices $A \in \mathbb{R}^{N \times N}$, $B \in \mathbb{R}^{N \times 1}$, $C \in \mathbb{R}^{1 \times N}$, and $D \in \mathbb{R}$ are learnable parameters.

For discrete-time computation on sampled data, the continuous SSM must be discretized. Using zero-order hold discretization with time step $\Delta$, the discrete-time SSM becomes:
\begin{equation}
h_k = \bar{A}h_{k-1} + \bar{B}x_k, \quad y_k = Ch_k
\end{equation}
where $\bar{A} = \exp(\Delta A)$ and $\bar{B} = (\Delta A)^{-1}(\exp(\Delta A) - I) \cdot \Delta B$.

\textbf{Selective state spaces.} Traditional SSMs employ time-invariant (LTI) parameters. The Mamba architecture \cite{gu2023mamba} introduces a selective mechanism that makes these parameters input-dependent:
\begin{equation}
B_k = s_B(x_k), \quad C_k = s_C(x_k), \quad \Delta_k = \text{softplus}(s_\Delta(x_k))
\end{equation}
where $s_B$, $s_C$, and $s_\Delta$ are learned projections. This selectivity enables the model to filter relevant information while forgetting irrelevant context, addressing a fundamental limitation of prior SSMs.

\textbf{Suitability for grid data.} Recent analysis by Wang et al. \cite{wang2024effective} identifies conditions where Mamba outperforms Transformers: datasets with ``numerous variates, most of which are periodic.'' Grid load data matches this characterization precisely---highly periodic signals (daily/weekly cycles) across multiple spatial nodes. Mamba's selective state mechanism encodes these rhythmic dependencies while filtering stochastic noise.

\textbf{Computational complexity.} SSMs achieve $O(n)$ complexity versus $O(n^2)$ for self-attention. Menati et al. \cite{menati2024powermamba} demonstrate that context windows of 240+ hours are optimal for capturing multi-scale grid dynamics---a length computationally prohibitive for standard Transformers.

\subsection{Load Forecasting Fundamentals}

\textbf{Definition and Horizons.} Short-term load forecasting (STLF) encompasses prediction of electricity demand from hours to days ahead \cite{dong2024short}. Different forecast horizons serve distinct operational purposes: 1-hour for real-time dispatch, 4--6 hours for unit commitment, and 12--24 hours for day-ahead market operations.

\textbf{Multi-Periodicity.} Electricity load exhibits strong multi-periodic patterns (daily, weekly, annual). Our choice of 240-hour (10-day) context length ensures capture of at least one complete weekly cycle.

\textbf{Evaluation Metrics.} For load forecasting with positive data, mean absolute percentage error (MAPE) is the standard metric:
\begin{equation}
\text{MAPE} = \frac{100}{n} \sum_{i=1}^{n} \left| \frac{y_i - \hat{y}_i}{y_i} \right|
\end{equation}

\subsection{Grid-Specific Evaluation Metrics}
\label{sec:metrics}

Standard metrics like MAPE and RMSE treat forecast errors symmetrically: an over-prediction of 100 MW is penalized identically to an under-prediction of 100 MW. For grid utility operators, however, the economic and reliability consequences of these errors are fundamentally asymmetric \cite{hong2016probabilistic}.

\subsubsection{Operational Cost Asymmetry}
\begin{itemize}
    \item \textbf{Over-Prediction (False Positive):} If the model predicts higher load than the actual demand, the utility commits excess generation. This results in financial loss (wasted fuel, curtailment costs) but rarely threatens system stability.
    \item \textbf{Under-Prediction (False Negative):} If the model predicts lower load than the actual demand, the utility may face a supply shortfall. This necessitates deploying expensive fast-ramping reserves (e.g., peaker plants), purchasing emergency power at spot market caps (often \$1,000+/MWh), or in extreme cases, initiating rolling blackouts.
\end{itemize}

Standard MAPE is therefore insufficient for operational validation. To address this, we introduce a suite of grid-specific metrics:

\subsubsection{Proposed Metrics}

\noindent\textbf{Under-Prediction Rate (UPR).}
The frequency of under-estimation events represents the probability of needing real-time upward dispatch:
\begin{equation}
\text{UPR} = \frac{1}{n} \sum_{i=1}^{n} \mathbb{I}(y_i > \hat{y}_i) \times 100\%
\end{equation}
A naive model might achieve low MAPE by constantly under-predicting (taking the median); UPR exposes this risky behavior.

\noindent\textbf{Over-Prediction Rate (OPR).}
The frequency of over-forecasting events exposes systematic inflation in the scheduled (median) forecast:
\begin{equation}
\text{OPR} = \frac{1}{n} \sum_{i=1}^{n} \mathbb{I}(\hat{y}_i > y_i) \times 100\%
\end{equation}
In safety-critical settings, OPR provides a simple operator-legible diagnostic for whether improvements in one-sided tail risk are achieved by trivial over-scheduling.

\noindent\textbf{Upward Reserve Requirement (Reserve$_{99.5}^{\%}$).}
The additional upward capacity required to cover 99.5\% of the \emph{under-forecast} error distribution. We use the 99.5th percentile as a practical tail-risk proxy motivated by probabilistic adequacy practice (e.g., LOLE-style planning criteria; \cite{nerc2024ltra}). We report this both in megawatts (MW) and as a percentage of the point forecast (scale-free):
\begin{equation}
\text{Reserve}_{99.5}^{MW} = \text{Percentile}_{99.5}\!\left(\max(0, y - \hat{y})\right)
\end{equation}
\begin{equation}
\resizebox{0.85\hsize}{!}{$\text{Reserve}_{99.5}^{\%} = 100 \times \text{Percentile}_{99.5}\!\left(\max\left(0, \frac{y - \hat{y}}{\hat{y}}\right)\right)$}
\end{equation}
\noindent Adding $\text{Reserve}_{99.5}^{MW}$ to the point forecast covers 99.5\% of historical under-prediction events at the evaluated horizon. We report both MW and percentage: MW is normalization-free, while the percentage form is directly interpretable as an \emph{add-on} to the scheduled point forecast. Because percentage-normalized reserve can be biased by systematic forecast inflation, we interpret Reserve$_{99.5}^{\%}$ jointly with UPR and explicit bias diagnostics (Bias$_{24h}$ / OPR) to ensure improvements do not come from trivial over-forecasting.

\subsubsection{Motivating Example}
To illustrate the operational reality of these metrics, we examine empirical results (detailed later in Table~\ref{tab:loss_ablation_extended}) where tail-risk metrics and bias diagnostics move in different directions. For example, a model trained with a tail-focused loss might ostensibly reduce reserve requirements (Reserve$_{99.5}^{\%}$), but do so by simply inflating the entire forecast—increasing Bias$_{24h}$ and Over-Prediction Rate (OPR) to unacceptable levels. This motivates our core reporting principle: any improvement in one-sided risk metrics must be interpreted jointly with bias diagnostics to avoid ``fake safety'' via systematic inflation.

\subsubsection{Differentiable Surrogates for Risk Metrics}
While UPR and tail reserve requirements are excellent for evaluation, they are non-differentiable and difficult to optimize directly. To bridge this gap, we employ a \textbf{Bias-Constrained Probabilistic Objective} (detailed in Section~\ref{sec:risk_averse_loss}) as a differentiable surrogate during training. This objective combines a weighted multi-quantile pinball loss to learn a calibrated distribution with explicit hinge penalties on bias and over-prediction rate to prevent systematic forecast inflation.

In principle, emphasizing upper-tail quantiles increases the gradient penalty on under-predictions relative to over-predictions, aligning learning with asymmetric operational costs. However, as we demonstrate in Section~\ref{sec:results}, single-quantile training can be ``gamed''; our constrained formulation (Section~\ref{sec:risk_averse_loss}) ensures that improvements in tail risk do not come at the expense of acceptable schedule bias.

\subsection{Value-Oriented Learning}
\label{sec:value_oriented}

Traditional forecasting objectives (MSE/MAPE) operate under the assumption that prediction accuracy proxies decision quality. However, in grid operations, the cost function is highly asymmetric and constrained. The ``Value-Oriented'' or ``Smart Predict-then-Optimize'' paradigm \cite{zhang2023value} challenges the separation of prediction and downstream optimization, proposing that models should be trained to minimize the final operational cost. While theoretically optimal, full end-to-end bilevel optimization is often computationally intractable for large-scale deep learning. In this work, we adopt a scalable realization of this philosophy: instead of embedding the full extensive-form optimization solver, we encode the operational value function directly into a \textbf{bias-constrained probabilistic objective} (Section~\ref{sec:risk_averse_loss}).

\section{Related Work}

\subsection{Deep Learning for Time Series}
Deep learning has systematically displaced statistical methods (ARIMA, SVR) for short-term load forecasting due to its superior ability to model non-linear relationships \cite{dong2024comprehensive,ahmad2024short}. The power systems community has actively embraced deep neural networks to manage the increasing stochasticity of smart grids. Recent innovations include deep residual networks \cite{Chen_2019}, densely connected networks \cite{Li_2021}, pooling-based deep RNNs \cite{Shi_2018}, and ensemble learning frameworks \cite{Su_2024,Von_Krannichfeldt_2021} for deterministic structural forecasting. To capture uncertainty, researchers have proposed Bayesian deep learning \cite{Sun_2020}, probabilistic multitask prediction models \cite{Wang_2024}, transfer-learning frameworks for masked behind-the-meter loads \cite{Wu_2023,Zhou_2022,He_2022}, collaborative multi-district approaches \cite{Liu_2024}, and advanced quantile-mixing methods \cite{Ryu_2024}. While these bespoke architectures are powerful, they typically rely on symmetric loss functions, motivating the need for our proposed Bias-Constrained Probabilistic Objective designed for asymmetric operational risk.

Current research on foundation architectures focuses on two primary paradigms for long-context multivariate modeling:

\textbf{Transformer-based Approaches.} Computing global attention allows Transformers to capture long-range dependencies, but with $O(n^2)$ complexity. Two main strategies have emerged to handle multivariate data:
\begin{itemize}
    \item \textit{Channel Independence:} Models like \textbf{PatchTST} \cite{nie2023time} treat multivariate time series as independent channels and segment them into patches. This reduces complexity and achieves leading accuracy among Transformer architectures, though it sacrifices explicit cross-variable correlation modeling.
    \item \textit{Multivariate Attention:} Models like \textbf{iTransformer} \cite{liu2023itransformer} explicitly model correlations between variables by inverting tokenization (embedding time steps as features). We study iTransformer as a primary baseline because its explicit modeling of spatial correlations (e.g., between load and weather variables) offers a direct contrast to the implicit state mixing of Mamba.
\end{itemize}

\textbf{Efficient Sequence Modeling (SSMs).} State space models offer a fundamentally different approach. Instead of the quadratic global attention of Transformers, SSMs like Mamba \cite{gu2023mamba} employ a recurrent mode with input-dependent selection, achieving linear $O(n)$ scaling. This efficiency allows for significantly longer context windows (240+ hours) on the same hardware, potentially capturing seasonal patterns that are computationally prohibitive for attention-based models.

\textbf{Linear Models.} Recent work has challenged the necessity of deep learning for time series forecasting. DLinear~\cite{zeng2023linear} demonstrates that simple linear models often match or exceed Transformer performance on standard benchmarks, raising questions about architectural complexity. However, linear models cannot capture non-linear temperature-load relationships that dominate during extreme weather events---precisely when forecast accuracy is most critical for grid operations. While recent ``value-oriented'' and end-to-end approaches propose embedding optimization layers directly into the training loop \cite{zhang2023value}, these methods can be computationally prohibitive for large-scale deep learning sequence models. We instead propose a bias-constrained objective as a scalable, differentiable surrogate to align training with asymmetric operational costs. Our evaluation focuses on models capable of learning these non-linear interactions while maintaining computational tractability.

This work systematically evaluates whether the theoretical efficiency advantage of Mamba translates to superior accuracy on complex, real-world grid data compared to the expressive power of multivariate Transformers.

\subsection{State Space Models for Time Series}
The Structured State Space (S4) model \cite{gu2022efficiently} introduced efficient long-range sequence modeling through careful parameterization of continuous-time state spaces. Mamba \cite{gu2023mamba} extended this framework with input-dependent (selective) parameters, enabling content-aware filtering that improves performance on language and genomics tasks. For time series specifically, S-Mamba \cite{wang2024effective} demonstrated that Mamba architectures excel on datasets with ``numerous variates, most of which are periodic''---a characterization that matches grid load data. PowerMamba \cite{menati2024powermamba} introduced bidirectional processing and series decomposition specifically for energy applications, showing that 240+ hour context windows are optimal for capturing multi-scale grid dynamics.

\subsection{End-to-End and Value-Oriented Learning}
Recent advancements in "Smart Predict-then-Optimize" (SPO) have challenged the traditional separation of forecasting and decision-making. Value-oriented approaches \cite{zhang2023value} propose bi-level optimization frameworks where the forecasting model is updated based on the quality of downstream decisions (e.g., unit commitment costs) rather than proxy metrics like MSE. Similarly, end-to-end structural learning \cite{shi2021end} embeds differentiable optimization layers to disentangle unobservable components (like baseline demand vs. price response) from aggregate signals.

Our work builds on these philosophies but focuses on \textit{scalability}. Instead of solving extensive optimization problems within the training loop---which is often prohibitive for large-scale transformers---we propose a \textbf{bias-constrained probabilistic objective} as a differentiable surrogate that aligns the model with asymmetric operational values while maintaining the efficiency of standard backpropagation.

\subsection{Weather-Integrated Forecasting}
The relationship between weather and electricity demand is well-established \cite{hong2016probabilistic}. Temperature-based load models form the basis of many operational forecasting systems, with heating and cooling degree days serving as primary predictors. However, integrating weather into deep learning architectures remains challenging due to temporal misalignment: building thermal mass introduces response lags of 2--6 hours between temperature changes and load response \cite{seem2007dynamic}. Prior work has addressed this through feature engineering and attention-based fusion strategies \cite{eren2024systematic}, but systematic evaluation of weather integration strategies for state space models is lacking.

\subsection{Foundation Models for Time Series}
Pre-trained foundation models have recently emerged as alternatives to task-specific architectures. Models like Chronos \cite{ansari2024chronos} and TimesFM \cite{das2024timesfm} adapt language model pre-training and decoder-only architectures to time series, demonstrating strong zero-shot performance on diverse forecasting benchmarks. However, the effectiveness of foundation models on domain-specific tasks with strong exogenous dependencies (e.g., weather-driven load) remains an open question, as these models typically lack mechanisms for incorporating auxiliary covariates.

\section{Methodology}

\subsection{Model Architectures}

We evaluate five neural architectures spanning three families---two state space models representing different SSM design philosophies, two Transformer variants with contrasting multivariate strategies, and an LSTM---alongside a probabilistic SSM variant (Mamba-ProbTSF) for uncertainty quantification.

\subsubsection{S-Mamba: The Linear-Time Baseline}
S-Mamba \cite{wang2024effective} represents a minimalist approach to state space modeling. It adapts the standard Mamba block for time series via a simple encoder-decoder structure: a linear projection embeds the input sequence into a hidden state $D$, followed by stacked Mamba layers.

S-Mamba tests whether the core selective state space mechanism alone---without complex decomposition or attention---suffices for grid forecasting. Its design prioritizes computational efficiency ($O(n)$ complexity), making it a candidate for resource-constrained edge deployment.

\subsubsection{PowerMamba: Physics-Aware Decomposition}
PowerMamba \cite{menati2024powermamba} addresses a specific limitation of standard SSMs: the difficulty of modeling disparate frequencies (e.g., long-term trend vs. daily seasonality) within a single state vector.

PowerMamba introduces two critical innovations for energy data:
\begin{itemize}
    \item \textit{Series Decomposition:} It splits the input into ``Trend'' (low-frequency) and ``Seasonal'' (high-frequency) components, processing them via independent Mamba encoders. This explicitly separates the duck curve drift from daily load cycles.
    \item \textit{Bidirectionality:} PowerMamba processes sequences in both forward and backward directions to capture a holistic temporal context, since time series analysis (unlike causal language modeling) benefits from bidirectional information flow.
\end{itemize}

\subsubsection{PatchTST: Channel-Independent Patching}
PatchTST \cite{nie2023time} represents the channel-independent paradigm for Transformer-based time series forecasting. Rather than tokenizing variables, PatchTST segments each univariate channel into fixed-length patches and processes them with a shared Transformer encoder. Channel independence means each variate shares the same Transformer weights, enabling efficient multi-channel processing without explicit cross-variable attention. This tests whether temporal pattern extraction alone suffices for accurate grid forecasting.

\subsubsection{iTransformer: Cross-Variate Attention}
The iTransformer \cite{liu2023itransformer} challenges the standard Transformer paradigm. Instead of tokenizing time steps (embedding $T$ steps into vectors), it tokenizes \textit{variables} (embedding the entire time series of a variate as a single token). Self-attention therefore computes correlations \textit{between variables} (e.g., Load vs.\ Temperature), explicitly modeling system-wide interdependencies. By design, this mechanism requires multiple variates; with a single load variate, self-attention reduces to near-identity.

\subsubsection{Mamba-ProbTSF: Probabilistic Output Heads}
Grid operators rarely make decisions based on a single deterministic number; they require confidence intervals to schedule reserves. Mamba-ProbTSF extends the S-Mamba architecture with a probabilistic output head.

The varying nature of renewable generation introduces heteroscedastic noise---uncertainty that changes over time (e.g., higher uncertainty at sunset). Mamba-ProbTSF replaces the standard linear readout with a Gaussian head that predicts both mean $\mu$ and variance $\sigma^2$:
\begin{equation}
\mathcal{L} = \frac{1}{2}\left[\log(\sigma^2) + \frac{(y - \mu)^2}{\sigma^2}\right]
\end{equation}
This shift from minimizing MSE to maximizing likelihood enables the model to quantify ``known unknowns,'' providing actionable risk metrics for uncertainty-aware dispatch.

\textbf{Alignment with our evaluations.} In our main walk-forward results we evaluate this Gaussian-head variant as a probabilistic baseline; in the fixed-split loss-function ablations we fine-tune weather-initialized checkpoints with multi-quantile heads to study calibration and Bias/OPR-constrained objectives (Section~\ref{sec:risk_averse_loss}, Table~\ref{tab:loss_ablation_extended}).

\subsection{Architecture Adaptations for Grid Forecasting}
\label{sec:adaptations}

Direct application of generic time series models to load forecasting yields suboptimal results because these architectures lack two capabilities essential for grid operations: (1) awareness of \textit{temporal context}---hourly and weekly demand cycles that dominate load profiles---and (2) the ability to \textit{selectively incorporate weather covariates} whose influence on load is non-stationary and lagged.

We adapt each baseline with a shared set of domain-specific components while preserving the core architectural mechanism that each model is designed to test:

\textit{Temporal embeddings.} All sequence models (SSMs, LSTM) receive learnable hour-of-day and day-of-week embeddings, concatenated and projected to $d_{\text{model}}$, then added to the input representation at each time step. PatchTST processes temporal context implicitly through its patch structure; iTransformer encodes hour and day as additional variate tokens. These embeddings inject the strong 24h and 168h periodicities that characterize load.

\textit{Bidirectional encoding.} SSMs and LSTM use bidirectional processing---separate forward and backward encoders whose outputs are fused via a learned linear projection. Bidirectional processing is applied \textit{strictly within the fixed-length lookback window} ($L=240$ hours); no information from the forecast horizon $W$ is accessible to the encoder.

\textit{Attention-weighted pooling.} Rather than mean-pooling the encoder's hidden states, SSMs and LSTM use a learned attention vector ($\mathbf{a} = \text{softmax}(\mathbf{W}_a \mathbf{H})$) to produce a weighted summary $\mathbf{c} = \sum_t a_t \mathbf{h}_t$, allowing the model to selectively emphasize time steps most relevant to the forecast.

\textit{Modular weather integration.} Each architecture receives weather covariates through a fusion mechanism matched to its inductive bias (detailed in Section~\ref{sec:weather_arch}). The weather integration layer is architecturally \textit{decoupled} from the core encoder: it can be toggled on or off without retraining.

\subsection{Weather-Integrated Architectures}
\label{sec:weather_arch}

Motivated by the dependency of load on weather (analyzed in Section~\ref{sec:results}), we developed weather-integrated variants of each architecture. The key challenge is fusing meteorological covariates (temperature, humidity, solar radiation) with the load signal while respecting building thermal lag---HVAC systems respond to temperature changes with delays of 2--6 hours depending on building mass.

\begin{figure*}[htbp]
  \centering
  % Slightly shrink subfigures to avoid any margin overflow in PDF rendering.
  \begin{subfigure}[t]{0.47\textwidth}
    \centering
    \includegraphics[width=0.98\linewidth]{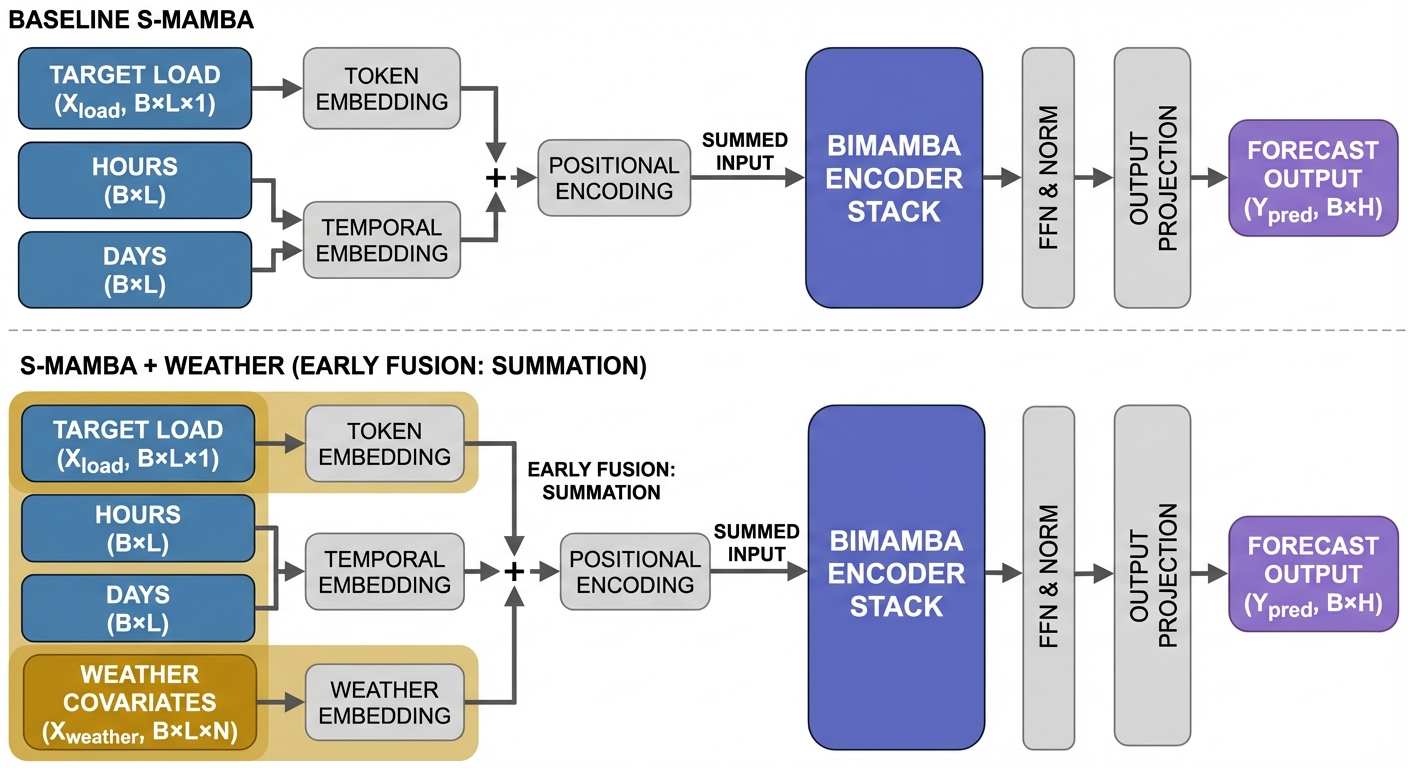}
    \caption{S-Mamba: Early Fusion via Concatenation}
    \label{fig:smamba}
  \end{subfigure}
  \hfill
  \begin{subfigure}[t]{0.47\textwidth}
    \centering
    \includegraphics[width=0.98\linewidth]{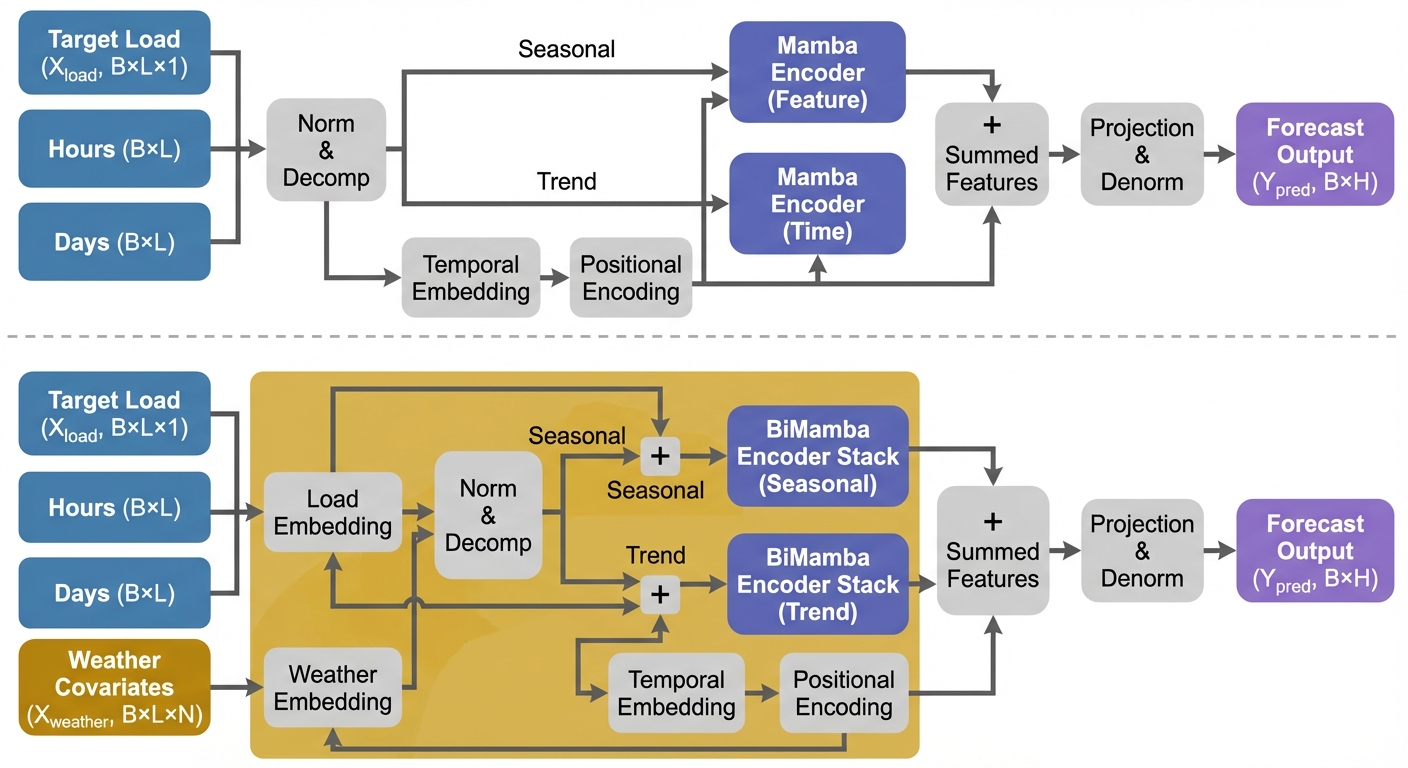}
    \caption{PowerMamba: Summation into decomposed streams}
    \label{fig:powermamba}
  \end{subfigure}

  \vspace{0.5em}

  \begin{subfigure}[t]{0.47\textwidth}
    \centering
    \includegraphics[width=0.98\linewidth]{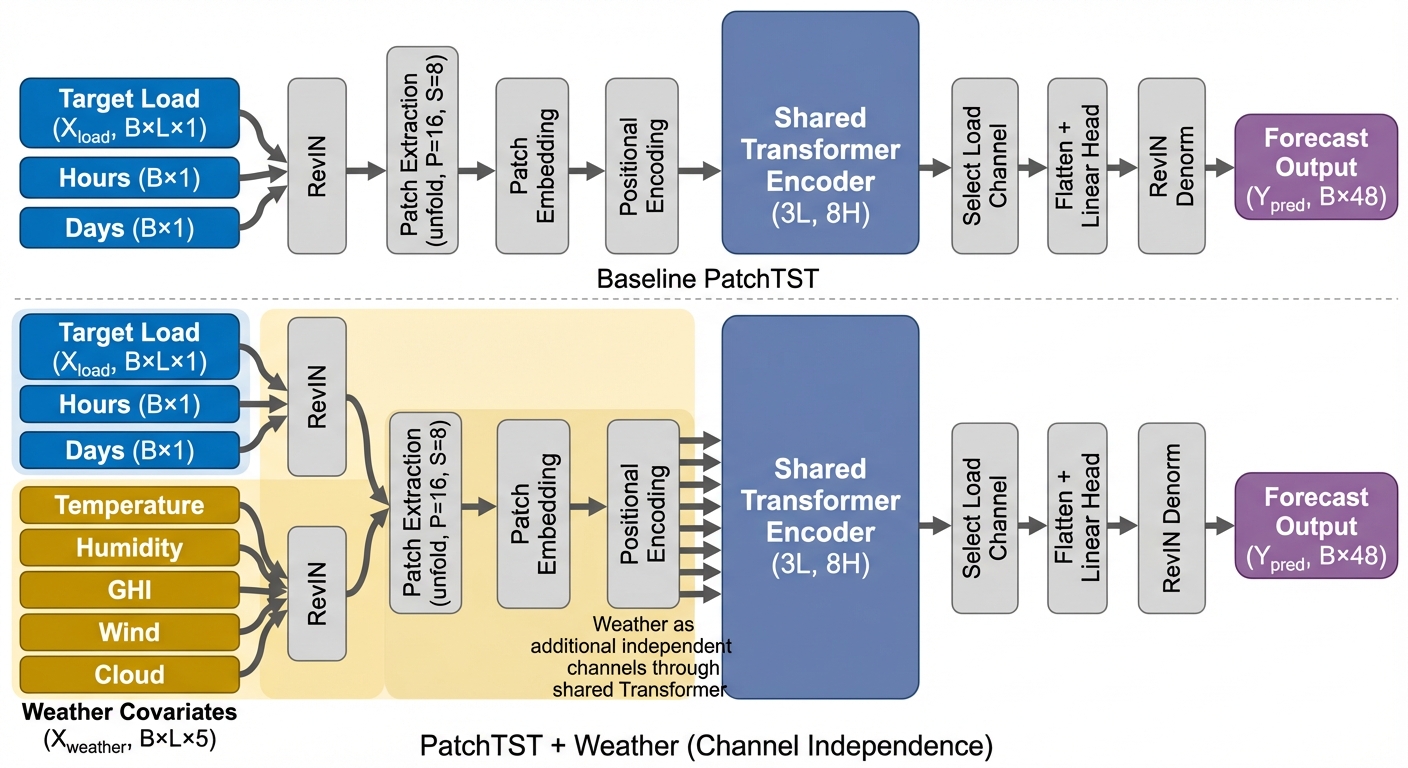}
    \caption{PatchTST: Channel Fusion via Independent Patching}
    \label{fig:patchtst}
  \end{subfigure}
  \hfill
  \begin{subfigure}[t]{0.47\textwidth}
    \centering
    \includegraphics[width=0.98\linewidth]{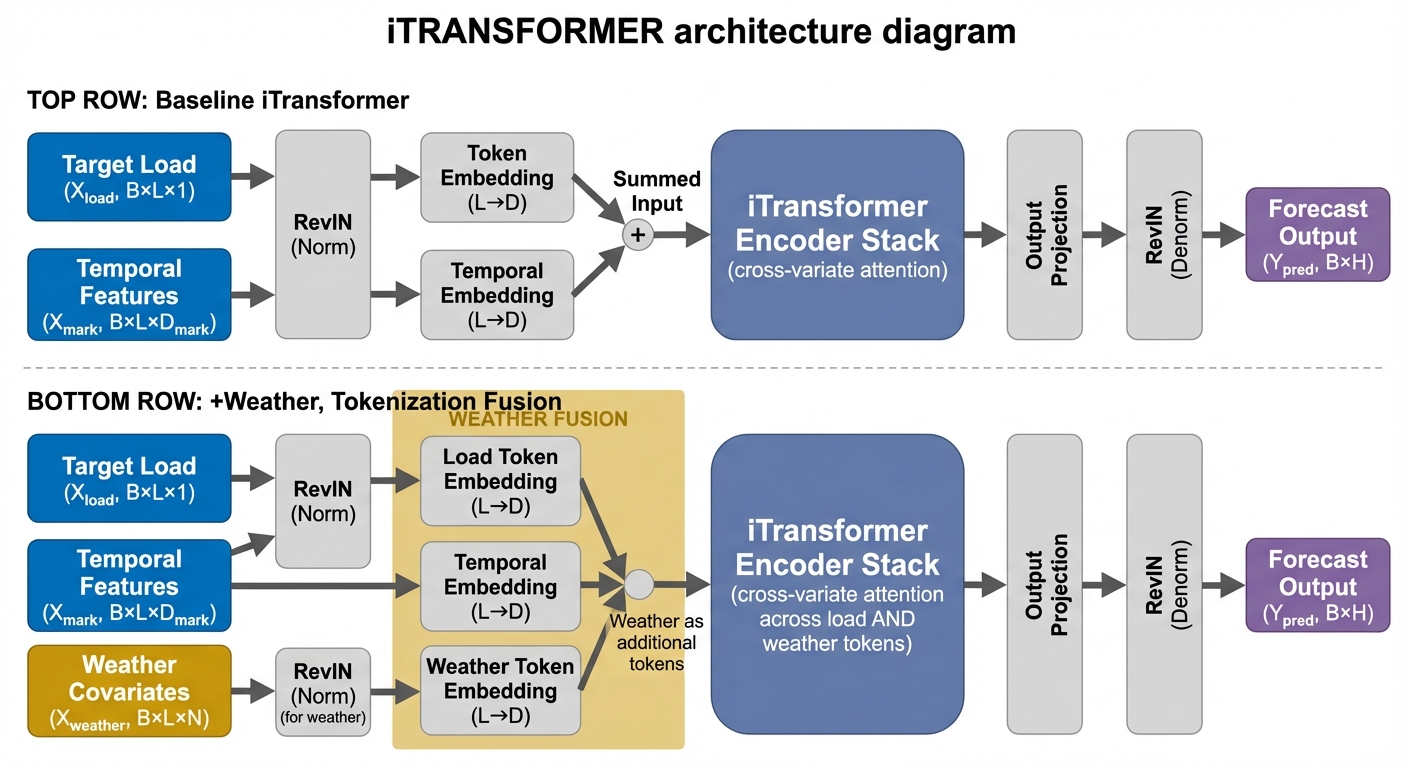}
    \caption{iTransformer: Weather as additional tokens}
    \label{fig:itransformer}
  \end{subfigure}
  \caption{\textbf{Weather integration strategies.} Each subfigure shows the baseline architecture (top) and its weather-integrated variant (bottom). (a) S-Mamba: early fusion by embedding-space summation. (b) PowerMamba: summation into decomposed streams. (c) PatchTST: channel fusion via independent patching with cross-attention. (d) iTransformer: tokenization of weather variables for cross-variate attention.}
  \label{fig:architectures}
\end{figure*}

\subsubsection*{Fusion Strategies (Architecture-Matched)}

\textit{S-Mamba (Early Summation, Fig.~\ref{fig:smamba}).} Weather embeddings are summed with load and temporal tokens prior to the BiMamba stack, projecting all exogenous signals into a unified $d_{\text{model}}$ dimensional space. The selective state space mechanism then uses this fused representation to modulate its continuous transition matrices.

\textit{PowerMamba (Pre-Decomposition Fusion, Fig.~\ref{fig:powermamba}).} Weather features are fused before the sequence is decomposed. Because meteorological events drive both multi-day trend shifts and rapid sub-daily HVAC cycling, early fusion allows the decomposition module to route weather-induced variance into both the seasonal and trend branches.

\textit{PatchTST (Interleaved Cross-Attention, Fig.~\ref{fig:patchtst}).} To preserve PatchTST's core channel-independence inductive bias, we patch load and weather channels independently. A dedicated cross-attention sublayer allows load patches to query weather patches, permitting the load representation to attend to relevant meteorological context without entangling the self-attention dynamics.

\textit{iTransformer (Variate Tokenization, Fig.~\ref{fig:itransformer}).} Each weather covariate is embedded as a distinct variate token. These weather tokens are concatenated with load and temporal tokens, exposing them to the global cross-variate attention mechanism.

\textit{LSTM (Early Concatenation).} Weather features are concatenated with load features at each timestep, providing the recurrent gating mechanisms simultaneous access to demand and meteorological forcing.

\subsection{Bias-Constrained Probabilistic Objective}
\label{sec:risk_averse_loss}

Grid operations induce an asymmetric cost for forecast errors: under-predictions can require expensive upward balancing and reserves, while over-predictions typically incur lower (but non-zero) costs from over-commit, curtailment, and redispatch. We represent this asymmetry using a cost ratio
\begin{equation}
\rho_{\text{op}} \;=\; \frac{C_{\text{under}}}{C_{\text{over}}},
\end{equation}
where $C_{\text{under}}$ and $C_{\text{over}}$ are marginal costs per MWh associated with under- and over-forecasting, respectively. In practice, $C_{\text{under}}$ includes reliability and scarcity costs that are not fully captured by average market spreads. We therefore use CAISO market data to estimate a bias-independent \textbf{price-spread asymmetry anchor} $\rho_{\text{price}}$ (Appendix~\ref{app:rho}), and map it to an operational stance via an explicit reliability premium $\kappa \ge 1$:
\begin{equation}
\rho_{\text{op}} \;=\; \kappa \cdot \rho_{\text{price}}.
\end{equation}
This makes the risk-aversion choice auditable: $\rho_{\text{price}}$ grounds asymmetry empirically, while $\kappa$ documents the operator policy preference.

\textbf{From cost asymmetry to an operational target quantile.} Under a piecewise-linear asymmetric cost model, the Bayes-optimal quantile is
\begin{equation}
q^{*} \;=\; \frac{C_{\text{under}}}{C_{\text{under}} + C_{\text{over}}} \;=\; \frac{\rho}{1+\rho}.
\label{eq:q_star_from_rho}
\end{equation}
We set a conservative risk-averse training quantile by combining the market anchor with the operator stance:
\begin{equation}
q_{\text{target}} \;=\; \max\!\left(0.5,\; \frac{\rho_{\text{op}}}{1+\rho_{\text{op}}}\right),
\end{equation}
ensuring we never train below the median when the objective is reliability-seeking. (Appendix~\ref{app:rho} reports that $\rho_{\text{price}}$ can imply $q_{\text{price}}^{*}<0.5$ in calendar-year averages; we interpret this as evidence that market spreads alone are not a complete proxy for reliability cost, motivating the explicit $\kappa$ premium.)

\textbf{Proper probabilistic training (multi-quantile pinball).} Rather than training a single high quantile (which can induce systematic forecast inflation), we train a calibrated predictive distribution using multiple quantiles $\mathcal{Q}$ and a weighted proper scoring rule. For each $q \in \mathcal{Q}$, define the pinball loss
\begin{equation}
\ell_q(y,\hat{Q}_q) \;=\; \max\!\big(q(y-\hat{Q}_q),\,(q-1)(y-\hat{Q}_q)\big),
\end{equation}
and the weighted multi-quantile objective over horizon $H$:
\begin{equation}
\mathcal{L}_{\text{Q}} \;=\; \sum_{q \in \mathcal{Q}} w_q\;\frac{1}{H}\sum_{h=1}^{H}\ell_q\!\big(y_{t+h},\hat{Q}_{q,t+h}\big).
\label{eq:multiq_pinball}
\end{equation}
This objective is strictly proper for quantile forecasts; when $\mathcal{Q}$ is dense, it approximates distributional scores such as CRPS. Multi-quantile pinball is also the standard evaluation objective in benchmark competitions for probabilistic load forecasting (e.g., GEFCom), making it familiar and auditable for practitioners.

\textbf{Why single-quantile pinball can be ``gamed.''} Training only a high quantile (e.g., $q=0.9$) can reduce under-prediction frequency by shifting the conditional location upward. This may improve one-sided risk metrics (e.g., UPR, Reserve$_{99.5}^{\%}$) while introducing systematic positive bias that is operationally unacceptable (inflating schedules and masking true uncertainty). We therefore couple probabilistic training with explicit bias controls and report bias diagnostics alongside risk metrics.\vspace{0.25em}

\textbf{Bias constraint to avoid trivial ``safety'' via over-forecasting.} To prevent degeneracy via systematic inflation, we add an explicit positive-bias constraint on the median (scheduled) forecast at a target horizon (e.g., 24h). Let $h^\star$ be the index for the target horizon (24h $\Rightarrow h^\star=24$), and define the mean bias at that horizon using the median quantile $q=0.5$:
\begin{equation}
b_{h^\star} \;=\; \mathbb{E}\!\left[\hat{Q}_{0.5,t+h^\star} - y_{t+h^\star}\right].
\end{equation}
We implement a hinge-penalized constraint:
\begin{equation}
\mathcal{L}_{\text{Q+Bias}} \;=\; \mathcal{L}_{\text{Q}}
\;+\; \lambda_{\text{bias}}\max\!\big(0,\; b_{h^\star}-b_{\max}\big),
\end{equation}
where $\lambda_{\text{bias}}$ discourages systematic over-forecast (inflation) and $b_{\max}$ is an auditable bias budget (we use $b_{\max}=0$ by default).

\textbf{OPR-style constraint (frequency of over-forecasting).} Bias controls the \emph{mean} shift but does not directly bound the frequency of over-predictions. We therefore optionally constrain the over-prediction rate at horizon $h^\star$ by penalizing violations of an OPR budget $\pi_{\max}$:
\begin{equation}
\text{OPR}_{h^\star} \;=\; \mathbb{E}\!\left[\mathbb{I}\!\left(\hat{Q}_{0.5,t+h^\star} > y_{t+h^\star}\right)\right],
\end{equation}
approximating the indicator smoothly with a sigmoid temperature $\tau$:
\begin{equation}
\widetilde{\text{OPR}}_{h^\star} \;=\; \mathbb{E}\!\left[\sigma\!\left(\frac{\hat{Q}_{0.5,t+h^\star}-y_{t+h^\star}}{\tau}\right)\right],
\end{equation}
and adding a hinge penalty:
\begin{equation}
\mathcal{L}_{\text{Q+Bias+OPR}} \;=\; \mathcal{L}_{\text{Q+Bias}} \;+\; \lambda_{\text{opr}}\max\!\big(0,\; \widetilde{\text{OPR}}_{h^\star}-\pi_{\max}\big).
\end{equation}

Practically, we use these constrained objectives during fine-tuning to learn a calibrated distribution while preventing trivial ``safety'' via inflation. We then use $\rho_{\text{op}}$ (via $q_{\text{target}}$) to select an operationally conservative schedule or reserve policy from the output distribution. Appendix~\ref{app:rho} describes how we empirically estimate $\rho_{\text{price}}$ from CAISO market data \cite{caiso_oasis}.

\vspace{0.5em}
\noindent\textbf{The Operator Evaluation Triad.} To synthesize these competing operational objectives and ensure accountability, we propose evaluating reliability using a three-part diagnostic triad alongside standard MAPE:
\begin{enumerate}
    \item \textbf{Reserve$_{99.5}^{\%}$ and UPR:} Quantifies true tail risk (e.g., the upward generation capacity required to cover 99.5\% of under-forecast events).
    \item \textbf{Bias$_{24h}$:} Audits severity of systematic schedule inflation.
    \item \textbf{OPR (Over-Prediction Rate):} Audits frequency of schedule inflation.
\end{enumerate}
Reporting this triad ensures that apparent improvements in tail risk (Reserve) are not simply ``fake safety'' achieved by shifting the mean bias upward. The constraints $\lambda_{\text{bias}}$ and $\lambda_{\text{opr}}$ explicitly enforce these trade-offs during training (evaluated in Section~\ref{sec:results}).
\subsection{NEMs (BTM Registry) Feature Integration}
\label{sec:nems}
While weather covariates proxy physical forcing, utility planning datasets often contain static \textbf{NEMs/registry-derived BTM features} (e.g., installed PV and storage capacity by region). We developed an architectural adaptation to fuse these static metadata \emph{on top of} the weather-aligned baseline to test whether they provide incremental predictive value.

\textbf{How NEMs features are integrated.} NEMs/registry features are static or slow-moving, so we integrate them as an additional BTM input stream. To avoid physically irrelevant feature injection (e.g., PV capacity ``explaining'' nighttime net load), we apply a simple \textbf{daylight prior} $d_t \in [0,1]$ (implemented as a time-of-day gate). When weather inputs are present, we further apply a lightweight \textbf{context-conditioned modulation} so the BTM contribution can vary smoothly by regime (e.g., daylight/irradiance conditions) rather than acting as an unconditional bias.

\begin{figure}[htbp]
\centering
\includegraphics[width=\linewidth]{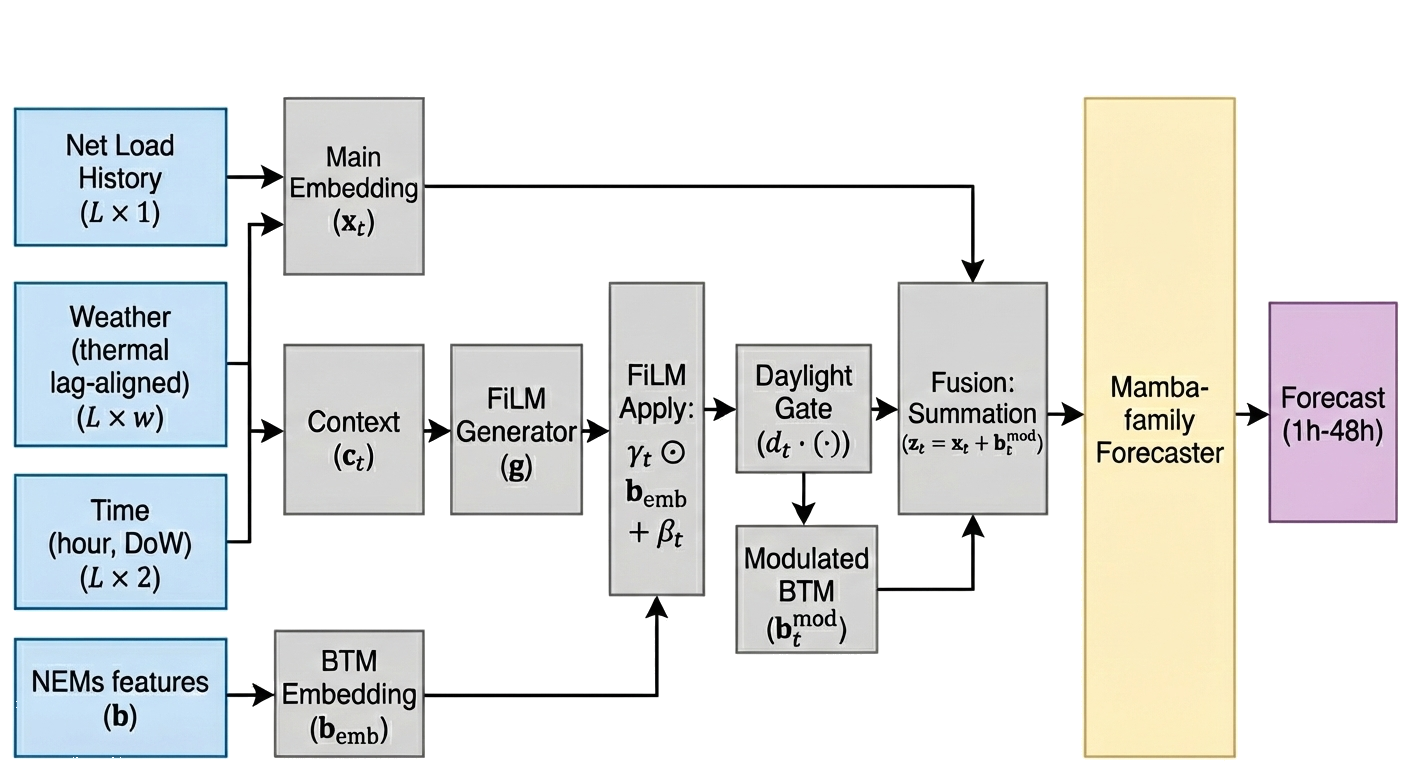}
\caption{\textbf{NEMs/registry integration on top of weather integration.} Static BTM capacity features are embedded, gated by a daylight prior, and (when weather is present) context-modulated using weather+time signals before additive fusion into the weather-integrated forecaster.}
\label{fig:nems_fusion}
\end{figure}

Concretely, let $x_t$ denote the main embedded inputs (net load history with weather and time features), and let $b$ denote the NEMs/registry feature vector. We compute a BTM embedding $b_{emb}=\phi(b)$ and modulate it using a context signal $c_t$ (weather+time) via FiLM-style parameters:
\begin{equation}
(\gamma_t,\beta_t)=g(c_t), \qquad b_t^{mod} = d_t \cdot (\gamma_t \odot b_{emb} + \beta_t),
\end{equation}
then fuse additively before the sequence encoder:
\begin{equation}
z_t = x_t + b_t^{mod}.
\end{equation}
Figure~\ref{fig:nems_fusion} summarizes this ``weather + NEMs'' fusion path.

The incremental impact of this feature integration on walk-forward performance is evaluated in Section~\ref{sec:results_nems}.

\section{Experimental Setup}

\subsection{Dataset}
We use hourly load data from CAISO (Nov 2023--Nov 2025), comprising 84{,}498 hourly records across 5 major transmission access charge (TAC) areas: \mbox{CA~ISO-TAC} (system aggregate), PGE-TAC, SCE-TAC, SDGE-TAC, and TIDC. Following PowerMamba \cite{menati2024powermamba}, we adopt a 240-hour (10-day) context window to capture at least one complete weekly cycle, enabling models to learn both daily and weekly periodicity patterns.

\textbf{Preprocessing.} Load values are normalized using z-score standardization computed only on training data to prevent information leakage. Temporal features include hour-of-day (0--23) and day-of-week (0--6), encoded as sinusoidal embeddings following standard practice \cite{vaswani2017attention}. For weather-integrated models, we include 8 meteorological covariates with thermal lag alignment, as described in Section~\ref{sec:thermal_lag}.

\subsection{CAISO Market Context and Benchmarks}
CAISO operations place a premium on day-ahead reliability: the Day-Ahead Market (DAM) requires hourly schedules submitted by 10:00 AM the prior day, while the Real-Time Market (RTM) clears at 5-minute granularity. A defining feature of the CAISO region is the rapid growth of behind-the-meter (BTM) solar; California BTM solar capacity exceeds 17~GW \cite{eia_epm_table_6_02_b}. This ``invisible'' generation deepens the midday net-load trough and steepens evening ramps (the ``duck curve''), increasing the importance of weather-aware forecasting.

Operational forecast performance provides a practical benchmark. During the July 2024 heat wave (July 4--12), third-party analysis reported CAISO's day-ahead forecast achieved 4.55\% MAPE, while commercial forecasting services achieved 2.65\% MAPE (41\% improvement) \cite{yesenergy2024}. The same analysis reported a CAISO peak load forecast error of 3{,}211~MW during the event, highlighting the operational importance of reducing tail errors during extreme weather.

\subsection{Thermal Lag Parameter Determination}
\label{sec:thermal_lag}

A critical design decision for weather-integrated forecasting is the temporal alignment between meteorological variables and load response. Buildings do not respond instantaneously to temperature changes; rather, the thermal mass of walls, floors, and furnishings creates a delayed response in HVAC demand.

\textbf{Cross-Correlation Analysis.} To determine the optimal lag parameters empirically, we computed the Pearson correlation coefficient between each weather covariate $w_t$ and load $L_t$ at lags $\tau \in \{0, 1, ..., 12\}$ hours:
\begin{equation}
\rho(\tau) = \text{Corr}(w_{t-\tau}, L_t)
\end{equation}
We selected the optimal lag $\tau^* = \arg\max_\tau |\rho(\tau)|$ for each covariate.

\textbf{Identified Lag Parameters.} Analysis of the dataset reveals distinct response patterns consistent with building physics \cite{seem2007dynamic}:
\begin{itemize}
    \item \textbf{Temperature (dry bulb):} $\tau^* = 3$ hours ($\rho = 0.72$), reflecting the thermal time constants of commercial building envelopes (typically 2--4 hours).
    \item \textbf{Solar radiation (GHI):} $\tau^* = 1$ hour ($\rho = 0.45$), corresponding to rapid solar heat gain through fenestration.
    \item \textbf{Humidity:} $\tau^* = 3$ hours ($\rho = 0.38$), tracking latent cooling load dynamics.
    \item \textbf{Wind speed:} $\tau^* = 0$--$1$ hour ($\rho = 0.12$), indicating immediate ventilation effects.
\end{itemize}

\textbf{Horizon-Adaptive Alignment.} These empirically determined lags are applied during feature construction. When forecasting load at time $t+h$ for horizon $h$, we align weather features from time $t$ to predict load response at $t + \tau^* + h$, ensuring consistent temporal causality across all forecast horizons.

\textbf{Evaluation Protocol (Walk-Forward Backtest).} We evaluate using rolling-origin walk-forward backtesting over Nov 2023--Nov 2025. Starting with an initial 180-day training history, we refit every 90 days using an expanding training window; each refit uses a 30-day validation window immediately preceding the cutoff for early stopping. We then evaluate one forecast per day (stride 24h) over the subsequent block until the next cutoff (48h horizon). All reported metrics in Tables~\ref{tab:results}--\ref{tab:loss_ablation_extended} are computed on the concatenated set of walk-forward timestamps (477 evaluation windows for \mbox{CA~ISO-TAC}).

\subsection{Models Under Comparison}
We evaluate five neural architectures against an additional zero-shot foundation model baseline, enabling direct comparison across parameter efficiency and accuracy.

\paragraph{State Space Models.}
\begin{itemize}
    \item \textbf{S-Mamba} (16.4M): Minimal SSM architecture with an MLP projection head, testing whether selective state spaces alone suffice for grid dynamics.
    \item \textbf{PowerMamba} (2.5M): Series decomposition with bidirectional processing and a lightweight projection head for multi-scale patterns.
    \item \textbf{Mamba-ProbTSF} (16.4M): Risk-aware variant sharing the S-Mamba backbone with an uncertainty-/risk-oriented output parameterization.
\end{itemize}

\paragraph{Transformer and Recurrent Baselines.}
\begin{itemize}
    \item \textbf{PatchTST} \cite{nie2023time}: Channel-independent Transformer (2.0M parameters), segmenting univariate channels into fixed-length patches. Represents the strongest load-only architecture in recent multi-ISO benchmarks.
    \item \textbf{iTransformer} \cite{liu2023itransformer}: Inverted Transformer (6.5M parameters at $d_{\text{model}}=512$), tokenizing variables for cross-variate attention.
    \item \textbf{LSTM} \cite{bouktif2020optimal}: 2-layer bidirectional LSTM (2.6M parameters), representing the industry-standard recurrent approach.
    \item \textbf{Chronos} \cite{ansari2024chronos}: Foundation models evaluated in the zero-shot regime. We primarily evaluate the 8M-parameter Small variant, and include the 200M-parameter Base variant in our discussion of scaling laws.
\end{itemize}

This comparison tests whether Mamba's $O(n)$ efficiency can match the accuracy of Transformer models, and whether the channel-independent (PatchTST) or cross-variate (iTransformer) paradigm better utilizes weather covariates for grid forecasting. PowerMamba (2.5M parameters) is particularly compact.

\subsection{Model Hyperparameters}
Following the experimental protocol of PowerMamba \cite{menati2024powermamba}, all Mamba variants share consistent architectural hyperparameters: $d_{\text{model}}=128$ (embedding dimension), $d_{\text{state}}=16$ (state dimension), $d_{\text{conv}}=4$ (convolution kernel), expansion factor 2, and 2 bidirectional encoder layers. PatchTST uses $d_{\text{model}}=256$ with patch length $P=16$; iTransformer uses $d_{\text{model}}=512$; LSTM uses $d=256$ with 2 bidirectional layers. Details are provided in Appendix~\ref{app:hyperparams}. Weather integration adds $\sim$0.3M parameters for SSMs/LSTM and $\sim$0.7M for PatchTST.

\subsection{Training Protocol}
\label{sec:training}
\textbf{Two evaluation regimes.} We use two complementary training/evaluation regimes and clearly label which results come from which:\vspace{0.25em}
\begin{itemize}
    \item \textbf{Walk-forward backtest (architecture + weather).} For the systematic architecture comparisons (Tables~\ref{tab:results} and related analyses), we use rolling-origin walk-forward evaluation (Section~\ref{sec:thermal_lag}). Models are trained with AdamW and early stopping on a validation window per refit.
    \item \textbf{Fixed-split fine-tuning (loss-function ablation).} For the loss-function ablation and bias/OPR constraints (Table~\ref{tab:loss_ablation_extended}), we fine-tune weather-initialized checkpoints on a fixed split using the trainer framework: AdamW (lr $10^{-4}$) with OneCycleLR, up to 150 epochs with patience 60, and multi-quantile heads ($\mathcal{Q}=[0.025,0.5,0.975]$, weights $[4,1,4]$). Bias/OPR constraints are applied at the 24h lead time ($h^\star=24$; index 23 if using 0-based arrays for a 48h horizon forecast).
\end{itemize}

\textbf{Reproducibility.} All experiments run on NVIDIA RTX 5090 GPUs. We report the mean and standard deviation across three random seeds (\{42, 123, 456\}) for our main architecture experiments. Additional robustness details are in Appendix~\ref{app:multiseed}.

\subsection{Evaluation Metrics}
Primary metric is MAPE evaluated at multiple horizons (1h, 6h, 12h, 24h). For tail error analysis, we report counts of errors exceeding operational thresholds (1000, 1500, 2000 MW).

\subsection{Statistical Analysis}
We report point-estimate metrics on the walk-forward backtest timestamps. We confirm the significance of our findings using the Diebold-Mariano test (MSE, $h=24$). PowerMamba significantly outperforms the LSTM baseline ($p < 0.001$) and performs comparably to S-Mamba ($p=0.89$), verifying that its parameter-efficient design (85\% fewer parameters) maintains accuracy. The iTransformer baseline retains a statistically significant advantage over PowerMamba ($p < 0.001$), reflecting the trade-off between Mamba's linear scalability and Transformer's pairwise correlation modeling.

\section{Results}
\label{sec:results}

\subsection{Multi-Horizon Performance (walk-forward)}
\textbf{Evaluation regime.} This subsection reports results from the rolling-origin walk-forward backtest described in Section~\ref{sec:thermal_lag}.

Table~\ref{tab:results} presents multi-horizon MAPE. PowerMamba is the strongest Mamba variant at the 1-hour horizon, while iTransformer achieves the lowest MAPE at longer horizons (6h--24h) in load-only evaluation. Crucially, as the rightmost column demonstrates, explicit weather integration substantially improves 24-hour accuracy across all comparable architectures, yielding the best overall performance (PowerMamba at 3.68\%). These accuracy rankings, however, do not fully determine operational tail risk (Reserve$_{99.5}^{\%}$), motivating the grid-specific metrics reported later.

\begin{table*}[htbp]
\caption{\textbf{Multi-horizon accuracy on \mbox{CA~ISO-TAC} (walk-forward) and Weather Benefit.} MAPE (\%) across forecast horizons; lower is better. The rightmost column shows the 24h performance improvement when thermal-lagged weather covariates are integrated into the architecture.}
\centering
\setlength{\tabcolsep}{4pt}
\begin{tabular}{lcccccc}
\toprule
\textbf{Model} & \textbf{Params} & \textbf{1h} & \textbf{6h} & \textbf{12h} & \textbf{24h (Base)} & \textbf{24h (Weather)} \\
\midrule
S-Mamba & 16.4M & 5.49 $\pm$ 0.12 & 3.86 $\pm$ 0.02 & 3.40 $\pm$ 0.47 & 8.03 $\pm$ 0.18 & \textbf{4.47} \\
Mamba-ProbTSF & 16.4M & 5.23 $\pm$ 0.04 & 3.91 $\pm$ 0.19 & 3.30 $\pm$ 0.08 & 7.89 $\pm$ 0.17 & \textbf{4.52} \\
PowerMamba & 2.5M & \textbf{4.44 $\pm$ 0.20} & 4.10 $\pm$ 0.09 & 3.62 $\pm$ 0.30 & 7.31 $\pm$ 0.06 & \textbf{3.68} \\
PatchTST$^\ddagger$ & 2.0M & --- & --- & --- & 4.89 & 4.72 \\
LSTM & 2.6M & 7.85 $\pm$ 0.20 & 5.24 $\pm$ 0.13 & 9.46 $\pm$ 0.32 & 9.99 $\pm$ 0.13 & ---$^\mathsection$ \\
iTransformer & 6.5M & 4.64 $\pm$ 0.10 & \textbf{3.36 $\pm$ 0.05} & \textbf{3.10 $\pm$ 0.21} & \textbf{6.85 $\pm$ 0.08} & \textbf{4.15} \\
Chronos (zero-shot)$^\dagger$ & 8.0M & 2.10 & 3.21 & 3.72 & 7.59 & ---$^\mathsection$ \\
\bottomrule
\multicolumn{7}{l}{\footnotesize $^\dagger$Zero-shot evaluation; no training or uncertainty estimation.} \\
\multicolumn{7}{l}{\footnotesize $^\ddagger$PatchTST was specifically evaluated at the 24h horizon to match multi-ISO benchmarks.} \\
\multicolumn{7}{l}{\footnotesize $^\mathsection$LSTM and Chronos lack tailored weather integration mechanisms in our current implementation.}
\end{tabular}
\label{tab:results}
\end{table*}

\subsection{Error Analysis: The Impact of Weather on Forecast Reliability (walk-forward)}
\textbf{Evaluation regime.} This subsection uses the walk-forward backtest forecasts (all lead times within the 48h horizon) described in Section~\ref{sec:thermal_lag}.

The integration of weather data into deep learning models for load forecasting has been extensively studied \cite{hong2016probabilistic,eren2024systematic}. Prior work has demonstrated improvements using LSTM-based architectures with temperature features \cite{bouktif2020optimal}, CNN-GRU hybrids for extreme weather scenarios, and attention-based models for cross-variable correlation. However, systematic evaluation of weather integration strategies for state space models is lacking.

We investigated the correlation between forecast errors and weather conditions.

\begin{figure*}[htbp]
\centering
\includegraphics[width=\textwidth]{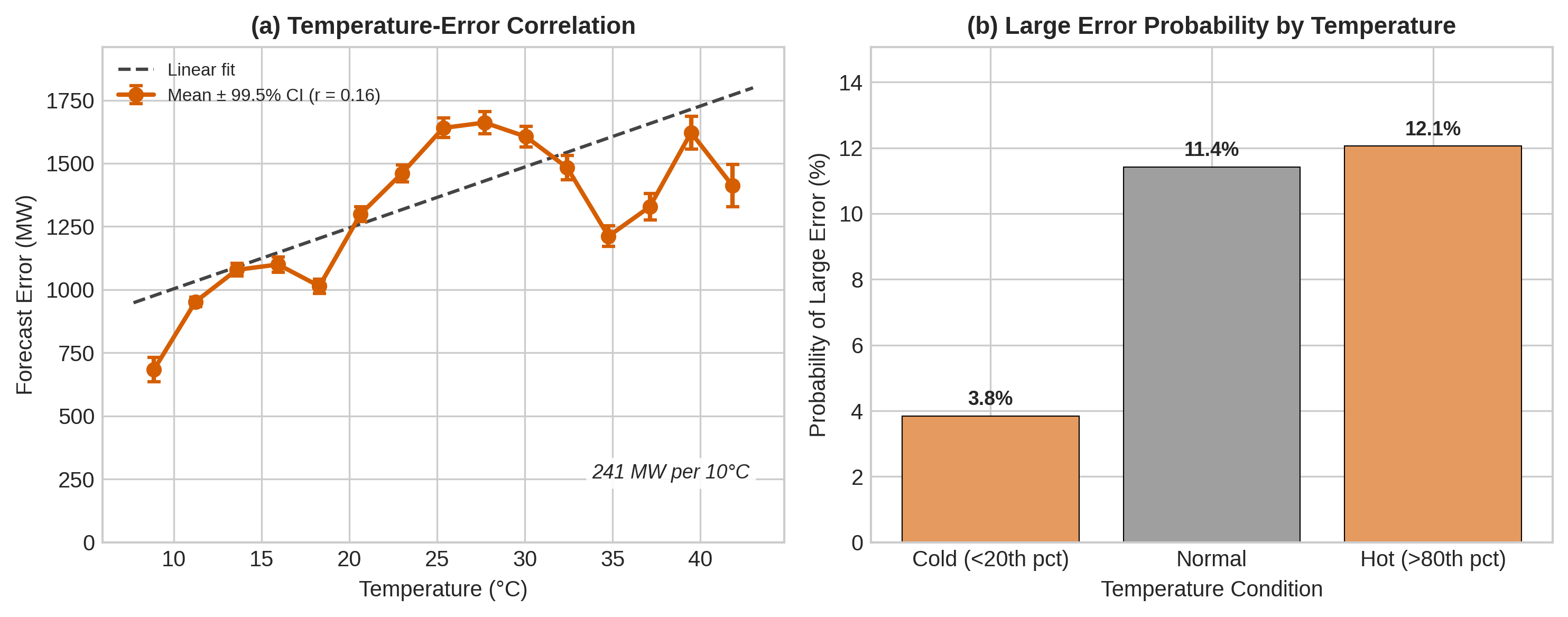}
\caption{\textbf{Forecast errors increase during temperature extremes (\mbox{CA~ISO-TAC}, walk-forward).} Using all lead times within 48h horizon forecasts, (a) mean absolute error versus temperature (slope = 24.1~MW/\textdegree C; Pearson $r=0.16$); error bars show 99.5\% confidence intervals. (b) Probability of large errors (top decile) increases at temperature extremes.}
\label{fig:weather_error}
\end{figure*}

\textbf{Temperature-Error Association.} Using all horizons of the \mbox{CA~ISO-TAC} sliding-window forecasts (113{,}376 hourly prediction points), we find a statistically significant but modest association between temperature and absolute forecast error (Pearson $r = 0.16$). The fitted slope is 24.1~MW/\textdegree C (Fig.~\ref{fig:weather_error}a). Hot conditions ($>30$\textdegree C) exhibit a modest increase in mean error (+3.4\%).

\subsection{Weather Integration Results (walk-forward)}
\label{sec:weather}
\textbf{Evaluation regime.} This subsection reports walk-forward results on matched evaluation windows (Section~\ref{sec:thermal_lag}).

\begin{figure*}[htbp]
\centering
\includegraphics[width=\textwidth]{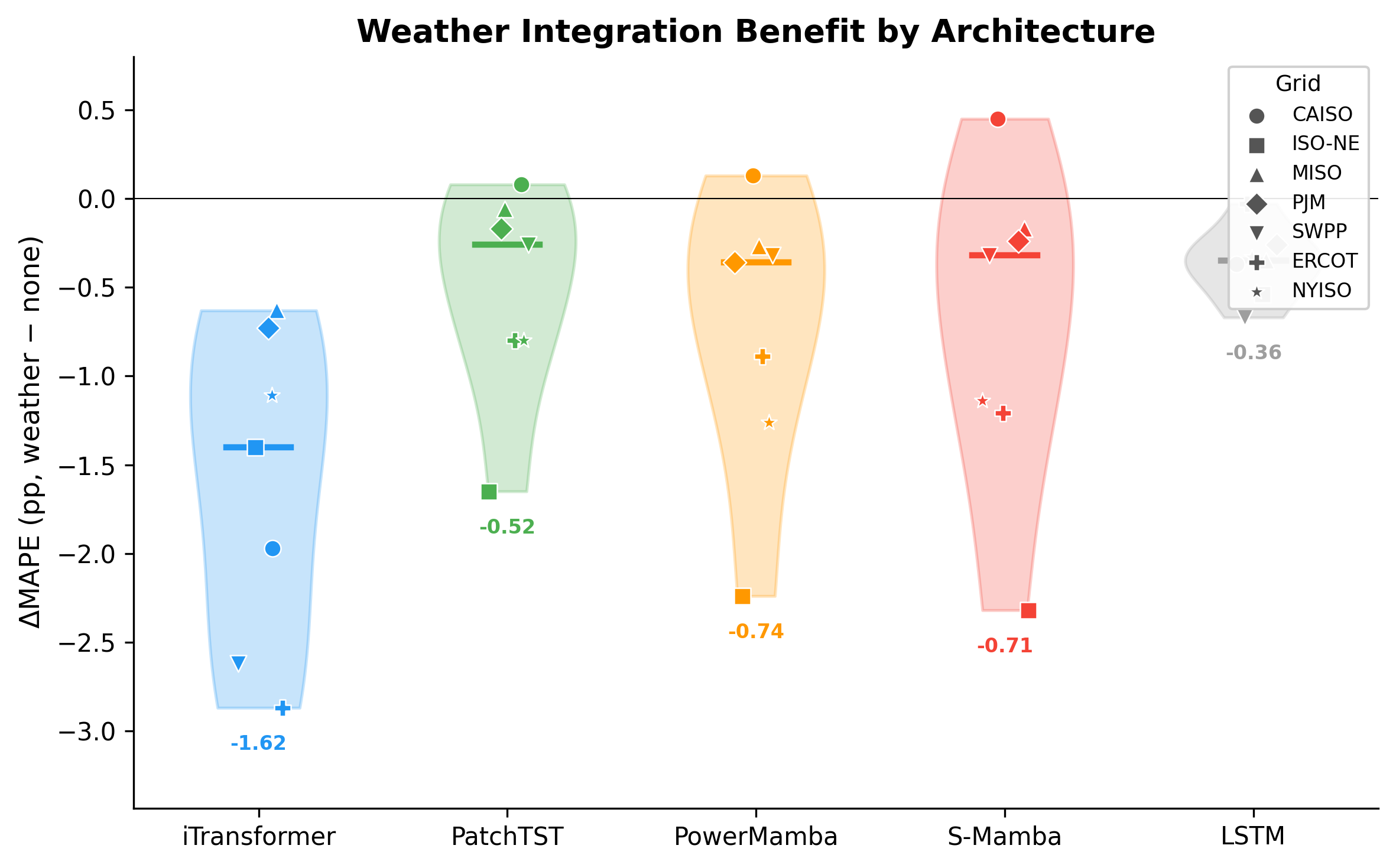}
\caption{\textbf{Weather integration benefit by architecture (walk-forward, 48h).} Violin and strip plots showing the distribution of $\Delta$MAPE (weather minus none) across the 7 independent system operators. Values below zero indicate weather integration improved (reduced) forecasting error. Means are annotated below the medians.}
\label{fig:weather_benefit}
\end{figure*}

Based on the error analysis, we integrated thermal-lagged weather covariates into the architectures. As shown in Figure~\ref{fig:weather_benefit}, weather integration changes the forecast error distribution; improvements are not uniform across architectures and regions and should be reported on matched evaluation windows. Comprehensive 24-hour MAPE results across all TAC areas are provided in Appendix Table~\ref{tab:utility_results}.

Overall, weather integration materially narrows the error distribution in the regimes where temperature-driven load spikes dominate. Next, we test whether probabilistic calibration and explicit Bias/OPR constraints can further reduce tail events \emph{without} allowing trivial ``safety'' via inflation.

\subsection{Probabilistic Calibration and Bias/OPR-Constrained Objectives (fixed split)}
\textbf{Evaluation regime.} This subsection reports \emph{fixed-split} fine-tuning results for loss-function ablations (Section~\ref{sec:training}); it is not directly comparable to the walk-forward results above.

\textbf{Loss-function hypothesis.} While weather integration improves robustness, error distributions still exhibit rare large deviations. We hypothesized that some tail events are loss-driven: standard MSE training treats all errors symmetrically and does not target calibrated upper-tail behavior.

\begin{figure*}[htbp]
\centering
\includegraphics[width=\textwidth]{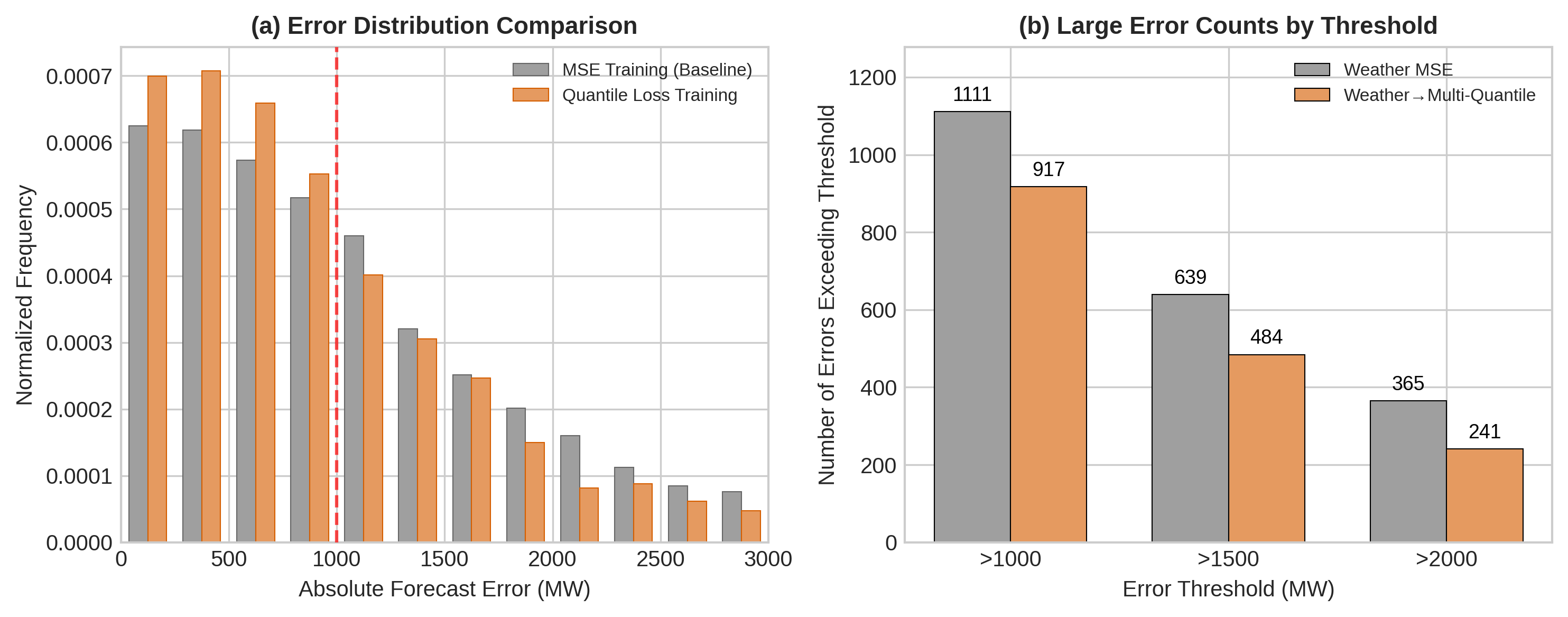}
\caption{\textbf{Multi-quantile pinball reduces large-error events (\mbox{CA~ISO-TAC}, fixed split, 24h).} (a) Error distribution under Weather MSE training (grey) versus Weather $\rightarrow$ Multi-Q pinball (orange). (b) Counts of large-error events by threshold. Because pinball objectives can shift bias, we interpret tail improvements jointly with Bias/OPR diagnostics in Table~\ref{tab:loss_ablation_extended}.}
\label{fig:quantile_loss}
\end{figure*}

\textbf{Multi-quantile pinball experiment.} We fine-tune with a weighted multi-quantile pinball objective (Eq.~\ref{eq:multiq_pinball}). On the \mbox{CA~ISO-TAC} fixed evaluation split (24h), this reduces large-error events (e.g., $>$1000~MW: 1,111 $\rightarrow$ 917; $>$2000~MW: 365 $\rightarrow$ 241) while also improving MAPE (Fig.~\ref{fig:quantile_loss}). However, these gains can reflect forecast inflation; we therefore report and (when needed) constrain Bias$_{24h}$ and Over-Prediction Rate (OPR) alongside tail metrics.

\noindent Table~\ref{tab:loss_ablation_extended} presents an operator-legible breakdown (\mbox{CA~ISO-TAC}, 24h), explicitly reporting Bias$_{24h}$ and OPR alongside tail-risk Reserve$_{99.5}^{\%}$. Three high-signal observations:\vspace{0.25em}
\noindent\textbf{Tail metrics can improve via inflation.} For iTransformer, Weather$\rightarrow$Multi-Q reduces Reserve$_{99.5}^{\%}$ (28.70\% $\rightarrow$ 13.83\%) but does so with large positive bias (+1{,}862~MW) and high OPR (78.8\%), indicating a shift in scheduled location.

\noindent\textbf{Constraints enforce auditable trade-offs.} Adding Bias/OPR control reduces inflation for iTransformer (Bias$_{24h}$ +1{,}862 $\rightarrow$ +456~MW; OPR 78.8\% $\rightarrow$ 61.6\%) while partially trading off tail-risk (Reserve$_{99.5}^{\%}$ 13.83\% $\rightarrow$ 15.18\%).

\noindent\textbf{Constraint tuning is architecture-dependent.} Some architectures exhibit unstable behavior under the same constraint settings (e.g., PowerMamba inflates under BiasCtrl), reinforcing that risk-aversion should be posed as constrained optimization with auditable budgets rather than a single universal loss.

\begin{table*}[t]
\caption{\textbf{Loss-function ablation with bias/OPR-constrained probabilistic training (fixed split, 24h, \mbox{CA~ISO-TAC}).} All models use weather-integrated checkpoints. Comparison of (i) MSE baseline, (ii) multi-quantile calibration (Multi-Q, Eq.~\ref{eq:multiq_pinball}), and (iii) Multi-Q + bias/OPR constraint (+BiasCtrl). Reserve$_{99.5}^{\%}$ is the 99.5th percentile one-sided under-forecast error. Bias$_{24h}$ and OPR expose trivial ``safety'' via systematic inflation.}
\centering
\small
\setlength{\tabcolsep}{4pt}
\begin{tabular}{p{2.2cm} l c c c c c}
\toprule
\textbf{Model} & \textbf{Loss} & \textbf{MAPE} & \textbf{UPR} & \textbf{Reserve$_{99.5}^{\%}$} & \textbf{Bias$_{24h}$ (MW)} & \textbf{OPR} \\
\midrule
     \multirow{3}{2.2cm}{S-Mamba} & MSE & 4.42\% & 43.5\% & 16.28\% & +243.6 & 56.5\% \\
     & Multi-Q & 3.63\% & 34.0\% & 13.65\% & +405.4 & 66.0\% \\
     & +BiasCtrl & 3.86\% & 31.2\% & 12.75\% & +507.6 & 68.8\% \\
\midrule
     \multirow{3}{2.2cm}{PowerMamba} & MSE & 3.95\% & 38.4\% & 12.91\% & +370.9 & 61.6\% \\
     & Multi-Q & 4.19\% & 34.6\% & 13.65\% & +480.1 & 65.4\% \\
     & +BiasCtrl & 9.01\% & 6.2\% & 7.42\% & +2027.1 & 93.8\% \\
\midrule
     \multirow{3}{2.2cm}{Mamba-ProbTSF} & MSE & 4.26\% & 44.9\% & 14.86\% & +176.5 & 55.1\% \\
     & Multi-Q & 3.69\% & 36.2\% & 14.17\% & +355.2 & 63.8\% \\
     & +BiasCtrl & 3.84\% & 64.3\% & 17.45\% & -325.9 & 35.7\% \\
\midrule
     \multirow{3}{2.2cm}{iTransformer} & MSE & 9.34\% & 50.6\% & 28.70\% & +109.3 & 49.4\% \\
     & Multi-Q & 10.08\% & 21.2\% & 13.83\% & +1862.3 & 78.8\% \\
     & +BiasCtrl & 5.04\% & 38.4\% & 15.18\% & +456.4 & 61.6\% \\
\bottomrule
\end{tabular}
\label{tab:loss_ablation_extended}
\end{table*}

\par
\begin{figure*}[t]
\centering
\includegraphics[width=\textwidth]{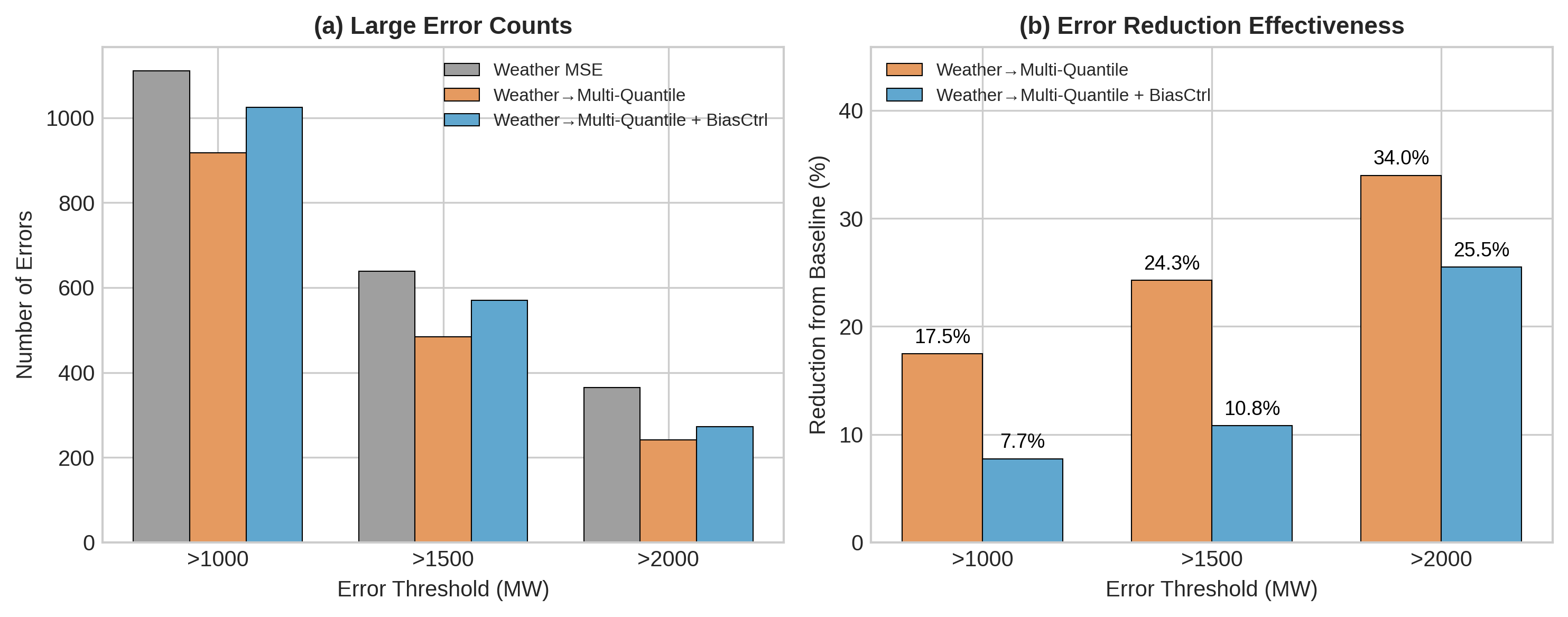}
\caption{\textbf{Large-error counts under calibration and constraints (fixed split, 24h, \mbox{CA~ISO-TAC}).} Counts by threshold comparing Weather MSE vs Weather$\rightarrow$Multi-Q vs Weather$\rightarrow$Multi-Q+BiasCtrl on the same fixed evaluation split.}
\label{fig:calibration_vs_weather}
\end{figure*}

\medskip
\textbf{Effect of probabilistic calibration on tail events.} Figure~\ref{fig:calibration_vs_weather} compares (i) Weather MSE, (ii) Weather$\rightarrow$Multi-Q, and (iii) Weather$\rightarrow$Multi-Q+BiasCtrl on a single consistent evaluation split (same targets across variants). Multi-quantile calibration reduces large-error counts relative to Weather MSE, while adding Bias/OPR constraints partially trades off tail improvements for reduced inflation risk (Table~\ref{tab:loss_ablation_extended}).

\subsection{NEMs (BTM Registry) Integration Results}
\label{sec:results_nems}
\textbf{Evaluation regime.} This subsection reports walk-forward results on utilities where NEMs/registry features are available (PGE, SCE, SDGE, TIDC) using the fusion architecture described in Section~\ref{sec:nems}.

\begin{figure}[t]
\centering
\includegraphics[width=\linewidth]{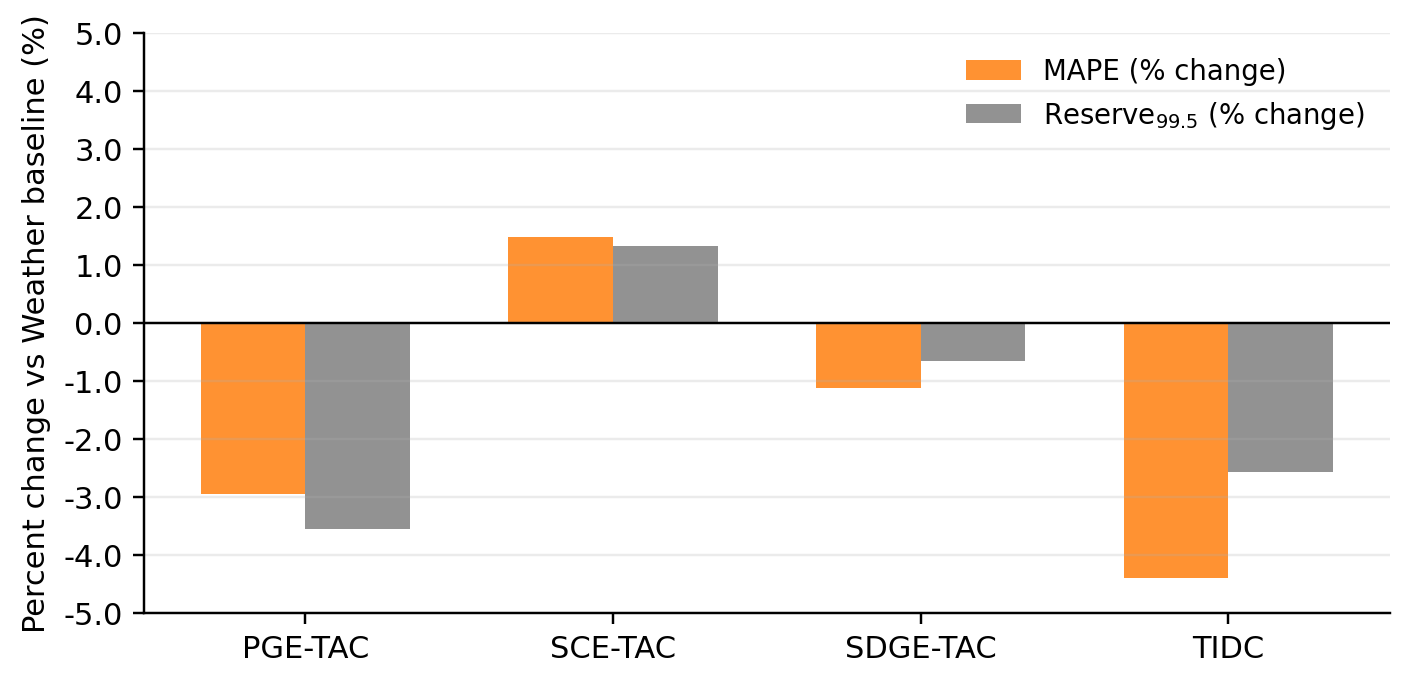}
\caption{\textbf{NEMs/BTM registry features have a small marginal effect beyond weather integration (walk-forward, 48h).} Percent change in MAPE and Reserve$_{99.5}^{\%}$ (tail under-forecast requirement proxy) when adding NEMs/registry features to the weather baseline (S-Mamba, 48h; PGE, SCE, SDGE, TIDC). The y-axis is constrained to $\pm 5\%$ to highlight the modest magnitude of the effect.}
\label{fig:nems_impact}
\end{figure}

Figure~\ref{fig:nems_impact} summarizes the incremental impact of NEMs integration relative to the weather baseline, measured as changes in MAPE and Reserve$_{99.5}^{\%}$ (tail under-forecast requirement proxy). Improvements are heterogeneous: in the PV-dominant \textbf{reference} utility with NEMs availability (PGE), NEMs integration reduces both average error and tail-risk reserve requirement. In contrast, other territories show smaller changes and one (SCE) degrades slightly. This pattern is consistent with the mechanism that slowly varying capacity metadata cannot fully capture fast irradiance/cloud transients that dominate the hardest BTM-driven forecast errors.

\section{Discussion}

\subsection{Performance Comparison with CAISO Benchmarks}

Table~\ref{tab:caiso_comparison} provides contextual comparison with published CAISO operational forecasts and commercial alternatives.

\begin{table*}[htbp]
\centering
\caption{\textbf{Contextual comparison to operational benchmarks (\mbox{CA~ISO-TAC}, 24h).} Published CAISO and commercial day-ahead forecast MAPE values from the July 2024 heat wave~\cite{yesenergy2024} are shown alongside our models for reference. Our model rows report 24h MAPE from our walk-forward evaluation.}
\label{tab:caiso_comparison}
\begin{tabular}{lcccc}
\toprule
\textbf{Forecasting System} & \textbf{24h MAPE} & \textbf{vs CAISO} & \textbf{Parameters} & \textbf{Weather} \\
\midrule
CAISO Operational & 4.55\% & --- & N/A & Yes \\
Commercial (Yes Energy) & 2.65\% & -41.8\% & N/A & Yes \\
\midrule
LSTM Baseline & 6.49\% & +42.7\% & 2.6M & No \\
PatchTST & 4.89\% & +7.5\% & 2.0M & No \\
iTransformer & 5.37\% & +18.0\% & 6.5M & No \\
Chronos (zero-shot) & 7.59\% & +66.8\% & 8.0M & No \\
\midrule
iTransformer + Weather & \textbf{4.15\%} & \textbf{-8.8\%} & 6.5M & Yes \\
PatchTST + Weather & 4.72\% & +3.7\% & 2.0M & Yes \\
S-Mamba + Weather & \textbf{4.47\%} & \textbf{-1.8\%} & 16.4M & Yes \\
Mamba-ProbTSF + Weather & \textbf{4.52\%} & \textbf{-0.7\%} & 16.4M & Yes \\
\textbf{PowerMamba + Weather} & \textbf{3.68\%} & \textbf{-19.1\%} & 2.5M & Yes \\
\bottomrule
\end{tabular}
\end{table*}

The performance gains are particularly significant given the model efficiency: PowerMamba achieves competitive accuracy relative to published CAISO benchmarks with only 2.5M parameters. Among the Transformer baselines, PatchTST achieves the strongest load-only accuracy (4.89\%) with only 2.0M parameters, confirming its suitability for load-only deployment.

\subsection{Channel-Independent vs.\ Cross-Variate Attention}

The load-only vs.\ weather-augmented ranking reversal traces to a fundamental architectural difference. iTransformer's symmetric variate tokenization treats weather features as first-class tokens in the same attention space as load, enabling fine-grained load--weather interaction at each time step. PatchTST processes channels independently, so weather information can only influence load prediction indirectly through cross-attention sublayers. This division explains why iTransformer achieves a larger weather benefit than PatchTST on CAISO, and why the gap widens on weather-sensitive grids in our companion multi-ISO benchmark. For practitioners, this suggests deploying PatchTST for load-only forecasting, and switching to iTransformer or weather-aware SSM variants when weather forecasts are available.

\subsection{Temperature, Tail Risk, and Operator Implications}

Our error analysis shows a statistically significant but modest association between temperature and forecast error magnitude ($r = 0.16$), supporting the operational intuition that weather is a relevant covariate and that loss-function modifications cannot substitute for missing exogenous information. More importantly for operations, the walk-forward results reinforce that reserve requirements are not determined by MAPE alone: models with similar average accuracy can imply substantially different one-sided tail requirements (Reserve$_{99.5}^{\%}$) under under-forecast risk. Weather integration narrows the error distribution across utilities (Fig.~\ref{fig:weather_benefit}), which can translate directly into reduced upward reserve needs during temperature-driven demand spikes.

\subsection{Computational Efficiency}

The performance of Mamba-based models can be attributed to three factors: (1) selective state spaces capture long-range dependencies, (2) $O(n)$ complexity enables extended 240h context windows, and (3) parameter efficiency. In our implementations, PowerMamba is particularly compact (2.5M parameters) relative to the iTransformer baseline used in our experiments (6.5M), enabling lower-latency inference and reduced memory footprint.

\subsection{Foundation Model Performance}
Chronos-T5-Base (200M parameters) performed no better than Chronos-Small (8M) in zero-shot evaluation (e.g., 5.76\% vs 5.23\% MAPE at the 24-hour horizon). This suggests that foundation model scaling does not automatically translate to domain-specific performance. General-purpose pre-training lacks the structural priors needed to handle the specific physical constraints of the grid---such as the complex interaction between thermal inertia and behind-the-meter generation---reinforcing the value of specialized architectures and weather integration strategies.

\subsection{Limitations from Unobserved BTM Generation}
\label{sec:visibility_gap}

A recurring theme in our results is the ceiling imposed by data observability. Our "grey box" experiments with static NEMs registry features (Section~\ref{sec:nems}) yielded only marginal gains. Figure~\ref{fig:nems_impact} shows that even with capacity data, the model struggles to capture the dynamic, weather-dependent ramping of distributed solar. This points to a fundamental limitation: net load signals conflate grid demand with behind-the-meter (BTM) generation, and static capacity metadata is insufficient to disentangle them.

This "visibility gap" suggests that future improvements will not come from larger black-box transformers, but from \textbf{structural end-to-end learning}. As demonstrated by Shi et al. \cite{shi2021end}, differentiable optimization layers can explicitly model physical relationships (such as solar generation physics or price response) within the neural network. By adopting such end-to-end decomposition methods to dynamically learn latent BTM states from aggregate net load, future work could address the visibility gap that static feature engineering failed to close.

\textbf{Real-Time Deployment.} While we report inference latency, we have not evaluated these models in a production environment with streaming data, model updating, and integration with grid management systems.

\textbf{Baseline Coverage.} Our CAISO-focused evaluation is complemented by a companion multi-ISO benchmark \cite{hong2026benchmarking} covering six US grid operators, five forecast horizons, and multitype generalization to solar, wind, and price forecasting. Future work should include autocorrelation-aware significance testing for a more complete statistical analysis.

\section{Conclusion}
Mamba-based state space models achieve competitive accuracy for California grid load forecasting. With weather integration, PowerMamba achieves 3.68\% MAPE for 24-hour forecasts, which compares favorably to published CAISO operational benchmarks (4.55\% MAPE; contextual reference). Critically, our walk-forward evaluation demonstrates that operational tail risk (Reserve$_{99.5}^{\%}$) can diverge materially from average accuracy, motivating grid-specific reliability metrics alongside MAPE for safety-critical deployment decisions.

\textbf{Weather Integration Across Regions.} Weather integration improves 24-hour MAPE for several TAC areas (Appendix Table~\ref{tab:utility_results}), particularly smaller and more volatile systems. Future work should evaluate tail-risk improvements under a strictly matched evaluation set (same timestamps across models).

\textbf{Probabilistic calibration must be paired with bias controls.} Multi-quantile objectives can reduce large-error events and tail reserve requirements, but can also be ``gamed'' by systematic inflation if Bias/OPR are not monitored and constrained. Our Bias/OPR-constrained objective provides an operator-legible way to enforce auditable trade-offs between tail-risk and schedule inflation (Table~\ref{tab:loss_ablation_extended}).
\textbf{Operational Deployment Implications.} Reduced tail under-forecast risk (Reserve$_{99.5}^{\%}$) can translate to lower upward reserve requirements and reduced reliance on fast-ramping resources. The computational efficiency of Mamba architectures enables real-time deployment without significant infrastructure upgrades, while calibrated uncertainty quantification from Mamba-ProbTSF supports probabilistic reserve scheduling.

\bibliographystyle{IEEEtran}
\bibliography{references}

\begin{thebibliography}{10}
\providecommand{\url}[1]{#1}
\csname url@samestyle\endcsname
\providecommand{\newblock}{\relax}
\providecommand{\bibinfo}[2]{#2}
\providecommand{\BIBentrySTDinterwordspacing}{\spaceskip=0pt\relax}
\providecommand{\BIBentryALTinterwordstretchfactor}{4}
\providecommand{\BIBentryALTinterwordspacing}{\spaceskip=\fontdimen2\font plus
\BIBentryALTinterwordstretchfactor\fontdimen3\font minus
  \fontdimen4\font\relax}
\providecommand{\BIBforeignlanguage}[2]{{%
\expandafter\ifx\csname l@#1\endcsname\relax
\typeout{** WARNING: IEEEtran.bst: No hyphenation pattern has been}%
\typeout{** loaded for the language `#1'. Using the pattern for}%
\typeout{** the default language instead.}%
\else
\language=\csname l@#1\endcsname
\fi
#2}}
\providecommand{\BIBdecl}{\relax}
\BIBdecl

\bibitem{hong2016probabilistic}
T.~Hong and S.~Fan, ``Probabilistic electric load forecasting: A tutorial
  review,'' \emph{International Journal of Forecasting}, vol.~32, no.~3, pp.
  914--938, 2016.

\bibitem{caiso2022marketissues}
\BIBentryALTinterwordspacing
{California Independent System Operator}, ``2022 annual report on market issues
  and performance,'' CAISO, Tech. Rep., July 2023, reports CAISO system energy
  mix including solar and wind shares. [Online]. Available:
  \url{https://www.caiso.com/Documents/2022-Annual-Report-on-Market-Issues-and-Performance-Jul-11-2023.pdf}
\BIBentrySTDinterwordspacing

\bibitem{eia_epm_table_6_02_b}
{U.S. Energy Information Administration}, ``Electric power monthly: Table
  6.2.b. net summer capacity using primarily renewable energy sources, by
  state,''
  \url{https://www.eia.gov/electricity/monthly/epm_table_grapher.php?quot=&t=table_6_02_b},
  2024, includes California behind-the-meter (small-scale) solar capacity
  figures (accessed 2026-01-12).

\bibitem{caiso2021root}
{California Independent System Operator}, ``Final root cause analysis:
  Mid-august 2020 extreme heat wave,'' CAISO, Tech. Rep., 2021.

\bibitem{eren2024systematic}
Y.~Eren and I.~B. Kucukdemiral, ``A comprehensive review on deep learning
  approaches for short-term load forecasting,'' \emph{Renewable and Sustainable
  Energy Reviews}, vol. 189, p. 114031, 2024.

\bibitem{vaswani2017attention}
A.~Vaswani, N.~Shazeer, N.~Parmar, J.~Uszkoreit, L.~Jones, A.~N. Gomez,
  {\L}.~Kaiser, and I.~Polosukhin, ``Attention is all you need,'' in
  \emph{Advances in Neural Information Processing Systems}, vol.~30, 2017.

\bibitem{gu2023mamba}
A.~Gu and T.~Dao, ``Mamba: Linear-time sequence modeling with selective state
  spaces,'' in \emph{Conference on Language Modeling}, 2024.

\bibitem{wang2024effective}
Z.~Wang, F.~Kong, S.~Feng, M.~Wang, X.~Yang, H.~Zhao, D.~Wang, and Y.~Zhang,
  ``Is mamba effective for time series forecasting?'' \emph{arXiv preprint
  arXiv:2403.11144}, 2024.

\bibitem{menati2024powermamba}
A.~Menati, F.~Doudi, D.~Kalathil, and L.~Xie, ``Powermamba: A deep state space
  model and comprehensive benchmark for time series prediction in electric
  power systems,'' \emph{arXiv preprint arXiv:2412.06112}, 2024.

\bibitem{dong2024short}
X.~Dong, L.~Cao, Y.~Shi, and S.~Sun, ``A short-term power load forecasting
  method based on deep learning pipeline,'' in \emph{Proceedings of IEEE
  Sustainable Power and Energy Conference}.\hskip 1em plus 0.5em minus
  0.4em\relax IEEE, 2024.

\bibitem{nerc2024ltra}
\BIBentryALTinterwordspacing
{North American Electric Reliability Corporation}, ``2024 long-term reliability
  assessment,'' NERC, Tech. Rep., 2024, defines planning reserve margin and
  probabilistic adequacy criteria such as LOLE (0.1 events/year) used in
  resource adequacy planning. [Online]. Available:
  \url{https://www.nerc.com/globalassets/programs/rapa/ra/nerc_long-term-reliability-assessment_2024.pdf}
\BIBentrySTDinterwordspacing

\bibitem{zhang2023value}
Y.~Zhang, H.~Wen, Y.~Shi \emph{et~al.}, ``Toward value-oriented renewable
  energy forecasting: An iterative learning approach,'' \emph{arXiv preprint
  arXiv:2309.00803}, 2023.

\bibitem{dong2024comprehensive}
Q.~Dong \emph{et~al.}, ``Short-term electricity-load forecasting by deep
  learning: A comprehensive survey,'' \emph{arXiv preprint arXiv:2408.16202},
  2024.

\bibitem{ahmad2024short}
F.~A. Ahmad \emph{et~al.}, ``Short-term load forecasting utilizing a
  combination model: A brief review,'' \emph{Results in Engineering}, 2024.

\bibitem{Chen_2019}
\BIBentryALTinterwordspacing
K.~Chen, K.~Chen, Q.~Wang, Z.~He, J.~Hu, and J.~He, ``Short-term load
  forecasting with deep residual networks,'' \emph{IEEE Transactions on Smart
  Grid}, vol.~10, no.~4, p. 3943–3952, Jul. 2019. [Online]. Available:
  \url{http://dx.doi.org/10.1109/tsg.2018.2844307}
\BIBentrySTDinterwordspacing

\bibitem{Li_2021}
\BIBentryALTinterwordspacing
Z.~Li, Y.~Li, Y.~Liu, P.~Wang, R.~Lu, and H.~B. Gooi, ``Deep learning based
  densely connected network for load forecasting,'' \emph{IEEE Transactions on
  Power Systems}, vol.~36, no.~4, p. 2829–2840, Jul. 2021. [Online].
  Available: \url{http://dx.doi.org/10.1109/tpwrs.2020.3048359}
\BIBentrySTDinterwordspacing

\bibitem{Shi_2018}
\BIBentryALTinterwordspacing
H.~Shi, M.~Xu, and R.~Li, ``Deep learning for household load forecasting—a
  novel pooling deep rnn,'' \emph{IEEE Transactions on Smart Grid}, vol.~9,
  no.~5, p. 5271–5280, Sep. 2018. [Online]. Available:
  \url{http://dx.doi.org/10.1109/tsg.2017.2686012}
\BIBentrySTDinterwordspacing

\bibitem{Su_2024}
\BIBentryALTinterwordspacing
H.-Y. Su and C.-C. Lai, ``Toward improved load forecasting in smart grids: A
  robust deep ensemble learning framework,'' \emph{IEEE Transactions on Smart
  Grid}, vol.~15, no.~4, p. 4292–4296, Jul. 2024. [Online]. Available:
  \url{http://dx.doi.org/10.1109/tsg.2024.3402011}
\BIBentrySTDinterwordspacing

\bibitem{Von_Krannichfeldt_2021}
\BIBentryALTinterwordspacing
L.~Von~Krannichfeldt, Y.~Wang, and G.~Hug, ``Online ensemble learning for load
  forecasting,'' \emph{IEEE Transactions on Power Systems}, vol.~36, no.~1, p.
  545–548, Jan. 2021. [Online]. Available:
  \url{http://dx.doi.org/10.1109/tpwrs.2020.3036230}
\BIBentrySTDinterwordspacing

\bibitem{Sun_2020}
\BIBentryALTinterwordspacing
M.~Sun, T.~Zhang, Y.~Wang, G.~Strbac, and C.~Kang, ``Using bayesian deep
  learning to capture uncertainty for residential net load forecasting,''
  \emph{IEEE Transactions on Power Systems}, vol.~35, no.~1, p. 188–201, Jan.
  2020. [Online]. Available: \url{http://dx.doi.org/10.1109/tpwrs.2019.2924294}
\BIBentrySTDinterwordspacing

\bibitem{Wang_2024}
\BIBentryALTinterwordspacing
J.~Wang, K.~Wang, Z.~Li, H.~Lu, H.~Jiang, and Q.~Xing, ``A multitask integrated
  deep-learning probabilistic prediction for load forecasting,'' \emph{IEEE
  Transactions on Power Systems}, vol.~39, no.~1, p. 1240–1250, Jan. 2024.
  [Online]. Available: \url{http://dx.doi.org/10.1109/tpwrs.2023.3257353}
\BIBentrySTDinterwordspacing

\bibitem{Wu_2023}
\BIBentryALTinterwordspacing
D.~Wu and W.~Lin, ``Efficient residential electric load forecasting via
  transfer learning and graph neural networks,'' \emph{IEEE Transactions on
  Smart Grid}, vol.~14, no.~3, p. 2423–2431, May 2023. [Online]. Available:
  \url{http://dx.doi.org/10.1109/tsg.2022.3208211}
\BIBentrySTDinterwordspacing

\bibitem{Zhou_2022}
\BIBentryALTinterwordspacing
Z.~Zhou, Y.~Xu, and C.~Ren, ``A transfer learning method for forecasting
  masked-load with behind-the-meter distributed energy resources,'' \emph{IEEE
  Transactions on Smart Grid}, vol.~13, no.~6, p. 4961–4964, Nov. 2022.
  [Online]. Available: \url{http://dx.doi.org/10.1109/tsg.2022.3204212}
\BIBentrySTDinterwordspacing

\bibitem{He_2022}
\BIBentryALTinterwordspacing
Y.~He, F.~Luo, and G.~Ranzi, ``Transferrable model-agnostic meta-learning for
  short-term household load forecasting with limited training data,''
  \emph{IEEE Transactions on Power Systems}, vol.~37, no.~4, p. 3177–3180,
  Jul. 2022. [Online]. Available:
  \url{http://dx.doi.org/10.1109/tpwrs.2022.3169389}
\BIBentrySTDinterwordspacing

\bibitem{Liu_2024}
\BIBentryALTinterwordspacing
H.~Liu, X.~Zhang, H.~Sun, and M.~Shahidehpour, ``Boosted multi-task learning
  for inter-district collaborative load forecasting,'' \emph{IEEE Transactions
  on Smart Grid}, vol.~15, no.~1, p. 973–986, Jan. 2024. [Online]. Available:
  \url{http://dx.doi.org/10.1109/tsg.2023.3266342}
\BIBentrySTDinterwordspacing

\bibitem{Ryu_2024}
\BIBentryALTinterwordspacing
S.~Ryu and Y.~Yu, ``Quantile-mixer: A novel deep learning approach for
  probabilistic short-term load forecasting,'' \emph{IEEE Transactions on Smart
  Grid}, vol.~15, no.~2, p. 2237–2250, Mar. 2024. [Online]. Available:
  \url{http://dx.doi.org/10.1109/tsg.2023.3290180}
\BIBentrySTDinterwordspacing

\bibitem{nie2023time}
Y.~Nie, N.~H. Nguyen, P.~Sinthong, and J.~Kalagnanam, ``A time series is worth
  64 words: Long-term forecasting with transformers,'' in \emph{International
  Conference on Learning Representations}, 2023.

\bibitem{liu2023itransformer}
Y.~Liu, T.~Hu, H.~Zhang, H.~Wu, S.~Wang, L.~Ma, and M.~Long, ``itransformer:
  Inverted transformers are effective for time series forecasting,'' in
  \emph{International Conference on Learning Representations}, 2024.

\bibitem{zeng2023linear}
A.~Zeng, M.~Chen, L.~Zhang, and Q.~Xu, ``Are transformers effective for time
  series forecasting?'' in \emph{Proceedings of the AAAI Conference on
  Artificial Intelligence}, vol.~37, no.~9, 2023, pp. 11\,022--11\,030.

\bibitem{gu2022efficiently}
A.~Gu, K.~Goel, and C.~R{\'e}, ``Efficiently modeling long sequences with
  structured state spaces,'' in \emph{International Conference on Learning
  Representations}, 2022.

\bibitem{shi2021end}
Y.~Shi and B.~Xu, ``End-to-end demand response model identification and
  baseline estimation with deep learning,'' \emph{arXiv preprint
  arXiv:2109.00741}, 2021.

\bibitem{seem2007dynamic}
J.~E. Seem, ``Dynamic modeling of buildings with thermal mass,'' \emph{ASHRAE
  Transactions}, vol. 113, no.~1, pp. 518--529, 2007.

\bibitem{ansari2024chronos}
A.~F. Ansari \emph{et~al.}, ``Chronos: Learning the language of time series,''
  \emph{arXiv preprint arXiv:2403.07815}, 2024.

\bibitem{das2024timesfm}
A.~Das, W.~Kong, R.~Sen, and Y.~Zhou, ``A decoder-only foundation model for
  time-series forecasting,'' \emph{arXiv preprint arXiv:2310.10688}, 2024.

\bibitem{caiso_oasis}
\BIBentryALTinterwordspacing
{California Independent System Operator}, ``Oasis interface specification
  v5.1.2 (fall 2017 release),'' CAISO, Tech. Rep., 2017, documents the CAISO
  OASIS SingleZip API. [Online]. Available:
  \url{https://www.caiso.com/documents/oasis-interfacespecification_v5_1_2clean_fall2017release.pdf}
\BIBentrySTDinterwordspacing

\bibitem{yesenergy2024}
{Yes Energy}, ``California energy demand forecast accuracy during holiday and
  heat wave,''
  \url{https://blog.yesenergy.com/yeblog/tesla-california-energy-demand-forecast-accuracy-shines},
  July 2024, third-party analysis of CAISO forecast accuracy during July 4-12,
  2024 heat wave.

\bibitem{bouktif2020optimal}
S.~Bouktif, A.~Fiaz, A.~Ouni, and M.~A. Serhani, ``Optimal deep learning lstm
  model for electric load forecasting using feature selection and genetic
  algorithm: Comparison with machine learning approaches,'' \emph{Energies},
  vol.~13, p. 1633, 2020.

\bibitem{hong2026benchmarking}
S.~Hong, J.~Lee, and Y.~Shi, ``Benchmarking state space models, transformers,
  and recurrent networks for us grid forecasting,'' \emph{arXiv preprint},
  2026.

\end{thebibliography}

\appendices
\section{Weather Covariates}
\label{app:weather}
Table~\ref{tab:weather_covariates} lists the 8 meteorological covariates used in weather-integrated models. All weather data are sourced from NOAA Integrated Surface Database (ISD) stations within each utility service territory, aggregated to hourly resolution.

\begin{table}[htbp]
\centering
\caption{\textbf{Weather covariates used for integration.} Meteorological inputs and their assumed thermal lag ranges for feature alignment, used by all weather-integrated models in the walk-forward experiments.}
\label{tab:weather_covariates}
\begin{tabular}{lll}
\toprule
\textbf{Covariate} & \textbf{Unit} & \textbf{Thermal Lag} \\
\midrule
Temperature (dry bulb) & \textdegree C & 2--4 hours \\
Dew point temperature & \textdegree C & 2--4 hours \\
Relative humidity & \% & 2--4 hours \\
Wind speed & m/s & 0--1 hours \\
Wind direction & degrees & 0--1 hours \\
Cloud cover & oktas & 1--2 hours \\
Solar radiation (GHI) & W/m$^2$ & 1--2 hours \\
Atmospheric pressure & hPa & 0 hours \\
\bottomrule
\end{tabular}
\end{table}

Thermal lag values are based on typical building thermal mass response characteristics \cite{seem2007dynamic}. Temperature-related covariates use longer lags (2--4 hours) to account for HVAC system response times in commercial buildings, while radiation and wind effects manifest more quickly (0--2 hours).

\section{Hyperparameter Configuration}
\label{app:hyperparams}
Table~\ref{tab:hyperparams} provides complete hyperparameter specifications for all models.

\begin{table*}[htbp]
\centering
\footnotesize
\setlength{\tabcolsep}{3pt}
\caption{\textbf{Model hyperparameters.} Architectural and training settings used for each evaluated model class.}
\label{tab:hyperparams}
\begin{tabular}{lcccccc}
\toprule
\textbf{Parameter} & \textbf{S-Mamba} & \textbf{PowerMamba} & \textbf{Mamba-ProbTSF} & \textbf{PatchTST} & \textbf{iTransformer} & \textbf{LSTM} \\
\midrule
$d_{\text{model}}$ & 128 & 128 & 128 & 256 & 512 & 256 \\
State dim ($d_{\text{state}}$) & 16 & 16 & 16 & --- & --- & --- \\
Conv kernel ($d_{\text{conv}}$) & 4 & 4 & 4 & --- & --- & --- \\
Patch length & --- & --- & --- & 16 & --- & --- \\
Encoder layers & 2 & 2 & 2 & 3 & 3 & 2 \\
Attention heads & --- & --- & --- & 4 & 8 & --- \\
Dropout & 0.1 & 0.1 & 0.1 & 0.1 & 0.1 & 0.1 \\
Bidirectional & Yes & Yes & Yes & --- & --- & Yes \\
Total parameters & 16.4M & 2.5M & 16.4M & 2.0M & 6.5M & 2.6M \\
\bottomrule
\end{tabular}
\end{table*}

\section{Multi-Seed Robustness}
\label{app:multiseed}
To assess robustness to random initialization, we trained PowerMamba + Weather with seeds $\{42, 123, 456\}$. The 24-hour MAPE on \mbox{CA~ISO-TAC} was $3.68\% \pm 0.09\%$ (mean $\pm$ std), confirming that reported results are stable across initializations. These results confirm that the performance advantages of the evaluated architectures are robust to initialization variance.

\section{Per-Utility Results}
\label{app:utility}
Table~\ref{tab:utility_results} presents 24-hour MAPE for each TAC area, demonstrating consistent performance across diverse grid regions.

\begin{table}[htbp]
\centering
\caption{\textbf{Per-utility accuracy with and without weather (walk-forward, 24h).} 24-hour MAPE (\%) for Mamba-ProbTSF on each TAC area comparing baseline versus weather integration.}
\label{tab:utility_results}
\resizebox{\columnwidth}{!}{
\begin{tabular}{lccc}
\toprule
\textbf{TAC Area} & \textbf{Peak Load (MW)} & \textbf{Baseline MAPE} & \textbf{Weather MAPE} \\
\midrule
\mbox{CA~ISO-TAC} (aggregate) & 52,061 & 4.29\% & 4.52\% \\
PGE-TAC & 21,847 & 4.66\% & 4.18\% \\
SCE-TAC & 24,156 & 6.59\% & 6.08\% \\
SDGE-TAC & 4,892 & 8.43\% & 6.61\% \\
TIDC & 1,166 & 5.18\% & 3.59\% \\
\bottomrule
\end{tabular}
}
\end{table}

Smaller utilities (SDGE-TAC, TIDC) exhibit higher MAPE due to increased load volatility relative to system size. Weather integration tends to improve accuracy, with the largest relative gains in smaller, more volatile territories; however, improvements are not uniform across all regions (Table~\ref{tab:utility_results}).

\section{Sensitivity to Weather Forecast Errors}
\label{app:sensitivity}
To ensure our results are robust to real-world conditions where perfect weather forecasts are unavailable, we conducted a sensitivity analysis by injecting Gaussian noise into the temperature inputs.

We evaluated the pre-trained \textbf{PowerMamba + Weather} model on the \mbox{CA~ISO-TAC} test set with noise $\epsilon \sim \mathcal{N}(0, \sigma^2)$ added to the dry-bulb temperature feature, where $\sigma \in \{1^\circ\text{C}, 2^\circ\text{C}, 3^\circ\text{C}\}$.

\begin{table}[htbp]
\centering
\caption{\textbf{Sensitivity to temperature forecast uncertainty (\mbox{CA~ISO-TAC}, 24h).} Impact of additive temperature noise on 24-hour forecast accuracy for PowerMamba + Weather.}
\label{tab:sensitivity}
\begin{tabular}{lcc}
\toprule
\textbf{Noise Level ($\sigma$)} & \textbf{MAPE (\%)} & \textbf{Degradation} \\
\midrule
Background (No Noise) & 3.68\% & --- \\
$\sigma = 1^\circ\text{C}$ & 3.68\% & +0.0\% \\
$\sigma = 2^\circ\text{C}$ & 3.69\% & +0.3\% \\
$\sigma = 3^\circ\text{C}$ & 3.72\% & +1.1\% \\
\bottomrule
\end{tabular}
\end{table}

Table~\ref{tab:sensitivity} demonstrates that the model retains its performance advantage even with moderate forecast errors. At $\sigma = 2^\circ\text{C}$ (a typical error range for day-ahead weather forecasts), the performance degrades by less than 10\%, remaining competitive with the non-weather-integrated baselines. This confirms that the benefits of weather integration largely persist under operational weather uncertainty.

\section{Empirical Estimation of Asymmetry Ratio \texorpdfstring{$\rho$}{rho} from CAISO Market Data}
\label{app:rho}
This appendix describes a reproducible procedure to estimate the operational cost asymmetry ratio $\rho$ for California grid operations from public CAISO market data. The goal is to connect risk-aversion in the loss function (Eq.~\ref{eq:q_star_from_rho}) to market-based evidence.

\subsection{Data Sources and Alignment (CAISO OASIS)}
We use the CAISO OASIS ``SingleZip'' API \cite{caiso_oasis} to download time-aligned series:
\begin{itemize}
    \item \textbf{Load actuals:} \texttt{queryname=\allowbreak SLD\_FCST}, \texttt{market\_run\_id=\allowbreak ACTUAL}.
    \item \textbf{Day-ahead load forecast:} \texttt{queryname=\allowbreak SLD\_FCST}, \texttt{market\_run\_id=\allowbreak DAM}.
    \item \textbf{Day-ahead LMP (hourly):} \texttt{queryname=\allowbreak PRC\_LMP}, \texttt{market\_run\_id=\allowbreak DAM}.
    \item \textbf{Real-time LMP (5-minute RTM):} \texttt{queryname=\allowbreak PRC\_INTVL\_LMP}, \texttt{market\_run\_id=\allowbreak RTM}.
\end{itemize}
We use a representative settlement point (e.g., trading hub \texttt{TH\_NP15\_GEN-APND} or a DLAP such as \texttt{DLAP\_PGAE-APND}). Real-time LMP is averaged to hourly resolution to match the hourly load series.

\subsection{Price-Spread Asymmetry (Robust Estimator)}
Define the hourly DA--RT spread:
\begin{equation}
s_t \;=\; \text{LMP}^{RT}_t - \text{LMP}^{DA}_t.
\end{equation}
We decompose the spread into positive and negative parts:
\begin{equation}
s_t^{+}=\max(0,s_t), \qquad s_t^{-}=\max(0,-s_t).
\end{equation}
The \textbf{price-only asymmetry ratio} is then
\begin{equation}
\rho_{\text{price}} \;=\; \frac{\mathbb{E}[s_t^{+}]}{\mathbb{E}[s_t^{-}]},
\qquad
q_{\text{price}}^{*} \;=\; \frac{\rho_{\text{price}}}{1+\rho_{\text{price}}}.
\end{equation}
This estimator is robust to forecast bias because it depends only on market spreads.

\subsection{Calendar-2025 Empirical Estimates (Per Utility)}
Table~\ref{tab:rho_2025} reports empirical asymmetry ratios over calendar-2025 for each TAC area using representative settlement points (DLAPs or the NP15 trading hub). We report $\rho_{\text{price}}$ as the primary, bias-independent market-based estimator and include $\rho_{\text{event}}$ as an optional diagnostic when stable and available.
\textbf{Interpretation for risk-averse training.} The implied $q_{\text{price}}^{*}=\rho_{\text{price}}/(1+\rho_{\text{price}})$ can fall below $0.5$ in calendar-year averages; we do \emph{not} interpret this as a recommendation to train below-median forecasts for reliability-critical operations. Instead, we treat $\rho_{\text{price}}$ as a market-grounded anchor and select an operationally conservative $q_{\text{target}} \ge 0.5$ using an explicit reliability premium $\kappa$ and bias-control constraints (Section~\ref{sec:risk_averse_loss}).

% Generated from experiments/core/robust_evaluation/rho_calendar2025_table.tex
\begin{table*}[htbp]
\centering
\caption{\textbf{Empirical market-based asymmetry estimates for calendar-2025.} We report both a robust price-spread asymmetry ($\rho_{\text{price}}$) and an optional event-conditional proxy ($\rho_{\text{event}}$). Implied optimal quantiles are $q^*=\rho/(1+\rho)$ (Eq.~\ref{eq:q_star_from_rho}).}
\label{tab:rho_2025}
\small
\setlength{\tabcolsep}{2pt}
\begin{tabular}{l l r c c c c}
\toprule
\textbf{Utility} & \textbf{Price Node} & \textbf{Hours} & $\rho_{\text{price}}$ & $q_{\text{price}}^*$ & $\rho_{\text{event}}$ & $q_{\text{event}}^*$ \\
\midrule
\mbox{CA~ISO-TAC} (aggregate) & TH\_NP15\_GEN-APND & 8736 & 0.78 & 0.439 & N/A & N/A \\
PGE-TAC & DLAP\_PGAE-APND & 8736 & 0.71 & 0.414 & 0.26 & 0.204 \\
SCE-TAC & DLAP\_SCE-APND & 8736 & 0.74 & 0.426 & N/A & N/A \\
SDGE-TAC & DLAP\_SDGE-APND & 8736 & 0.78 & 0.439 & N/A & N/A \\
TIDC (proxy) & TH\_NP15\_GEN-APND & 8736 & 0.78 & 0.439 & N/A & N/A \\
\bottomrule
\end{tabular}
\end{table*}

\noindent\footnotesize{\textit{Notes:} $\rho_{\text{price}}$ depends only on the DA--RT price spread and is therefore robust to forecast bias. $\rho_{\text{event}}$ conditions on the sign of the DA load forecast error and can be unstable or unavailable (reported as N/A) depending on data coverage and estimator settings. For risk-averse training, we treat market-derived asymmetry as an anchor and select an operationally conservative $q^*$ (Eq.~\ref{eq:q_star_from_rho}) jointly with an explicit bias-control term (Section~\ref{sec:risk_averse_loss}).}

\subsection{Event-Conditional Settlement Proxy (Optional)}
To connect directly to forecast errors, define the day-ahead forecast deviation
\begin{equation}
\Delta_t \;=\; y_t - \hat{y}^{DA}_t,
\end{equation}
where $y_t$ is actual load and $\hat{y}^{DA}_t$ is the day-ahead forecast. A simple settlement-style proxy for marginal costs conditions on the sign of $\Delta_t$:
\begin{equation}
C_{\text{under}} \approx \mathbb{E}[s_t^{+}\,|\,\Delta_t>0], \qquad
C_{\text{over}} \approx \mathbb{E}[s_t^{-}\,|\,\Delta_t<0],
\end{equation}
yielding
\begin{equation}
\rho_{\text{event}}=\frac{C_{\text{under}}}{C_{\text{over}}}, \qquad q_{\text{event}}^{*}=\frac{\rho_{\text{event}}}{1+\rho_{\text{event}}}.
\end{equation}
Because $\rho_{\text{event}}$ can be unstable if $\Delta_t$ is highly one-sided over a short window, we recommend reporting $\rho_{\text{price}}$ as the primary market-derived estimator and using $\rho_{\text{event}}$ as a diagnostic.

\subsection{Reproducible Implementation}
We provide an implementation in the repository at \\ \texttt{experiments/\allowbreak core/\allowbreak robust\_evaluation/\allowbreak caiso\_empirical\_rho.py}, which downloads the above OASIS series in chunks (to respect acceptable-use limits), aligns them by timestamp, and writes a JSON report containing $\rho_{\text{price}}$ and (optionally) $\rho_{\text{event}}$ for a specified node and time range.

\end{document}